\documentclass[lettersize,journal]{IEEEtran}
\usepackage{amsmath,amsfonts}
\usepackage{algpseudocode}
\usepackage{algorithm}
\usepackage{array}
\usepackage{textcomp}
\usepackage{stfloats}
\usepackage{url}
\usepackage{verbatim}
\usepackage{graphicx}
\usepackage{cite}
\usepackage{subcaption}
\usepackage{hyperref}
\usepackage{booktabs}
\usepackage[table]{xcolor}
\usepackage{bm}
\begin{document}

\title{MagLearn 2: High-Fidelity, Saturation-Aware, and History-Efficient Sequence-to-sequence Modeling of Transient B–H Behaviour Under PWM Excitation}

\author{
Jingrong Yang,~\IEEEmembership{Student Member, IEEE},
Binyu Cui,~\IEEEmembership{Student Member, IEEE},
Yuming Huo,
Lizhong Zhang, 
Alfonso Martinez,~\IEEEmembership{Member, IEEE},
Song Liu,
and Jun Wang,~\IEEEmembership{Member, IEEE}%
\thanks{This work was presented in part at the 10th International Power
Electronics Conference (IPEC-Nagasaki 2026--ECCE Asia), Nagasaki, Japan,
May 31--June 4, 2026.
(Corresponding author: Jun Wang.)}%
\thanks{Jingrong Yang, Binyu Cui, Yuming Huo, Lizhong Zhang, Song Liu,
and Jun Wang are with the Electrical Energy Management Group,
School of Electrical, Electronic and Mechanical Engineering,
University of Bristol, Bristol BS8 1TR, U.K.
(e-mail: \href{mailto:jingrong.yang@bristol.ac.uk}{jingrong.yang@bristol.ac.uk};\href{mailto:bc16461@bristol.ac.uk}{bc16461@bristol.ac.uk};
\href{mailto:song.liu@bristol.ac.uk}{song.liu@bristol.ac.uk};\href{mailto:jun.wang@bristol.ac.uk}{jun.wang@bristol.ac.uk}).}%
\thanks{Alfonso Martinez is with W\"urth Elektronik, Germany
(e-mail: \href{mailto:alfonso.martinez@we-online.com}{alfonso.martinez@we-online.com}).}%
}

\maketitle

\begin{abstract}
Accurate transient $B$–$H$ modeling under pulse-width-modulated (PWM) excitation is challenging since the magnetic-field response depends on both the instantaneous flux density and the preceding magnetisation trajectory. Although sequence-to-sequence models enable waveform prediction, existing studies have not jointly addressed how much magnetic history is required, how saturation region should be handled, and how trained models can be adapted to unseen materials. To address these gaps, this paper presents MagLearn 2, a condition-aware, short-window LSTM-based sequence-to-sequence framework for reconstructing the future $H(t)$ response from the  historical $B(t)$ and $H(t)$ trajectories. A saturation-aware auxiliary module selectively replaces saturation-dominated predictions within the complete waveform forecast. In a representative saturated 3C90 case, the auxiliary module reduces the sequence-level relative RMSE from 67.84\% to 9.19\%. To understand how much the history samples contributes to the predictions, an occlusion-based sensitivity analysis is conducted to quantifies the contribution of magnetic history and reveals that that the information most relevant to prediction is concentrated near the forecast boundary. In the independent evaluation for MagNet Challenge II, the proposed framework’s best-performing model achieved the lowest reported average RMSE to date on the sequence-prediction task, at 2.2\%.
\end{abstract}

\begin{IEEEkeywords}
transient magnetic modeling, magnetic hysteresis, power magnetics, data-driven method, machine learning, transfer learning
\end{IEEEkeywords}

\section{Introduction}
Advances in semiconductor devices have enabled higher switching frequencies and power densities in power electronic converter designs, while imposing increasingly demanding requirements on magnetic components \cite{7931709}. These advances are particularly important for data-centre power supplies requiring fast dynamic response under rapidly changing computational loads \cite{9984210}. However, as semiconductor switching performance improves, magnetic component remain major contributors to converter volume, power loss, and thermal constraints. Their core and winding losses, temperature rise, and nonlinear dynamic behaviour can therefore limit further improvements in converter efficiency, reliability, and control performance \cite{4239129}. Conventional modeling methods of magnetics in power electronics has primarily focused on steady state, where $B$–$H$ behaviour is characterised over complete excitation cycles. However, transient operation requires the instantaneous magnetic response to be determined from only the available preceding trajectory. Under pulse-width-modulated (PWM) excitation, the flux-density waveform $B(t)$ generally contains switching-frequency components superimposed on a lower-frequency envelope, producing nested and potentially unclosed minor loops in the $B$–$H$ trajectory \cite{7919225}. Since magnetic hysteresis is path dependent, an identical instantaneous $B(t)$ value may correspond to different $H(t)$ values depending on the preceding state and recent trajectory. Time-domain simulation is widely used in the converter design to evaluate transient electrical behaviour and provide sizing information for magnetic components before hardware implementation \cite{biela2010towards}. For voltage-driven magnetic components, $B(t)$ is determined by the applied excitation, whereas $H(t)$ represents the corresponding material response and determines the magnetising behaviour. Therefore, precise converter-level simulation requires magnetic models capable of reproducing the complete time-resolved $H(t)$ response.

Manufacturer datasheets typically provide permeability characteristics under specified test conditions, while in practical engineering calculations, the permeability of magnetic core is usually assumed to be constant. Although useful for preliminary design, this approximation neglects the nonlinear permeability and history-dependent local response of magnetic materials, limiting its accuracy in reproducing PWM-induced minor loops and saturation-dominated transient responses. Physics-based hysteresis models, including the Preisach and Jiles–Atherton formulations \cite{preisach1935magnetische,jiles1986theory}, represent magnetic memory, but are fundamentally constrained by minor-loop assumptions and a low-dimensional state description, respectively. Consequently, parameter calibration alone cannot reproduce the frequency-dependent and partially traversed $B$--$H$ trajectories generated under arbitrary PWM excitation. These limitations motivate a more complex data-driven transient Modeling approach that can learn the nonlinear, multivariable relationship directly from measured $B$–$H$ trajectories and their associated operating conditions.






Neural networks are universal function approximators and are natural for learning high-dimensional and nonlinear magnetic relationships directly from experimental data, without requiring a complete analytical description of the underlying magnetisation mechanisms \cite{lecun2015deep}. Earlier data-driven studies demonstrated their ability to model magnetic-material characteristics from measured data \cite{li2020magnet,antonio2021effective}. Supported by the MagNet database \cite{li2022magnet} and MagNet Challenge I \cite{chen2024magnet}, subsequent approaches extended this capability across arbitrary periodic waveforms and operating conditions. The LSTM-based Seq2Seq neural network datasheet \cite{9829998} and its Transformer-based encoder--projector--decoder extension \cite{li2023predicting} reconstructed complete $B$--$H$ loops by mapping $B(t)$ waveform to the corresponding $H(t)$ waveform, with the resulting loops also enabling core-loss estimation. These developments were subsequently integrated into the MagNet-AI framework \cite{li2023magnet_ai}. HARDCORE similarly employed a residual dilated CNN with circular padding to reconstruct the steady-state $H(t)$ waveform before estimating core loss \cite{kirchgassner2024hardcore}. MagLearn 1 formulated core-loss prediction as a sequence-to-scalar (Seq2Sca) problem and introduced a two-stage transfer-learning strategy comprising pre-training on data-rich materials and fine-tuning on the target material \cite{zhang2024maglearn}. Despite their different formulations, these approaches are designed for full-cycle modeling, where the complete excitation waveform is available. In contrast, the present task requires forecasting the unknown future $H(t)$ response from a partially observed magnetic trajectory. Therefore, these formulations do not directly address the present transient prediction problem.

The availability of transient measurements through MagNetX database \cite{kwon2025magnetx} and MagNet Challenge II extended data-driven magnetic modeling to time-resolved prediction under non-periodic PWM excitation. The MagNetX foundation model adopts a sequence-to-scalar LSTM and constructs a future trajectory autoregressively by returning each predicted $H$ value to the model memory \cite{wang2025magnetx}. While effective for sequential generation, this autoregressive formulation may accumulate prediction errors over longer forecast horizons. RHINO-MAG \cite{vater2026rhino} propagates a compact GRU hidden state after an initial warm-up interval, with particular emphasis on parameter efficiency, whereas LSTM-DAO \cite{wang2026framework} combines recurrent feature extraction with direction-dependent asymmetric hysteresis operators. These studies establish the feasibility of data-driven transient magnetic modeling, but some aspects remain less systematically investigated, including quantitative identification of the informative magnetic-history region, reduced robustness in saturation-dominated regions, and cross-material adaptation under limited target-material data. These considerations motivate several complementary design questions addressed by the MagLearn 2 framework developed in this work, building on the successful MagLearn 1 concept \cite{zhang2024maglearn}.

Among recurrent networks, long short-term memory (LSTM) is adopted for its gated memory mechanism, which retains relevant historical information over extended sequences while suppressing less informative observations \cite{hochreiter1997long}. Building on this temporal representation, the sequence-to-sequence (Seq2Seq) formulation maps a historical input sequence to a future output sequence and permits the history and forecast intervals to have different lengths \cite{sutskever2014sequence}.The preliminary MagLearn 2 study \cite{11597291} adopted this formulation and demonstrated short-window transient forecasting and few-shot adaptation. Building on this framework, the present work develops a condition-aware, short-window LSTM-based Seq2Seq framework for transient $B$–$H$ modeling under PWM excitation. The framework systematically addresses physical memory requirement, waveform-level prediction, saturation robustness, computational efficiency, and adaptation across magnetic materials.

The main contributions are summarised as follows:

\begin{itemize}
    \item A condition-aware Seq2Seq formulation is developed to predict the multi-step transient  $H(t)$  response from the historical $B$--$H$ trajectory and the known future $B(t)$ excitation, while allowing independent history and forecast length. A lightweight saturation-aware auxiliary module is introduced to selectively correct the saturation-dominated predictions, while preserving the non-saturated baseline response. This addresses a critical gap missing in existing approaches, leading to high prediction error in saturation regions.
    \item A performance-driven transfer-learning procedure selects a compatible base model through target-specific validation and applies material-specific fine-tuning under limited target-material data. The mean and 95th-percentile relative RMSEs average 3.75\% and 8.75\% across ten base-model materials, and 4.69\% and 10.83\% across five unseen materials after fine-tuning, with three of the five target materials achieving top-three rankings in MagNet Challenge II.
    \item The contribution of magnetic history to transient prediction is characterised through history-length analysis and the first application of occlusion-based sensitivity analysis to data-driven transient $B$--$H$ prediction. The results reveal that the informative magnetic memory is predominantly concentrated near the prediction boundary, which is a missing insight in existing approaches. This finding shows that history-dependent transient magnetic prediction does not necessarily require fundamental-cycle-scale memory and provides a quantitative basis for short-window modeling and history length selection.
\end{itemize}

\section{Motivation for Data-Driven Transient Magnetic Modeling}
Magnetic permeability ($\mu$), as one of the key parameters of ferromagnetic
materials, characterizes the nonlinear relationship between magnetic flux
density $B$ and magnetic field strength $H$. Magnetic component manufacturers
typically specify three types of permeability in their datasheets: initial
permeability ($\mu_i$), reversible permeability ($\mu_{\mathrm{rev}}$), and
amplitude permeability ($\mu_a$). These parameters are normally used in static
modeling for either small-signal circuits or idealized closed $B$--$H$ loops.

However, in practical power-electronics applications, the excitation is
large-signal, and the $B$--$H$ trajectory contains nested, unclosed minor loops
due to the coexistence of the fundamental and switching harmonics. In such
cases, as discussed in \cite{wang2025magnetx}, the incremental permeability
($\Delta\mu$) is more suitable to capture the time-varying $B$--$H$ dynamics.
Over a discrete interval $[t_{k-1}, t_k]$, the incremental permeability is
defined as

\begin{equation}
\Delta\mu
=
\frac{\Delta B(t)}{\mu_0 \Delta H(t)}
=
\frac{B(t_k)-B(t_{k-1})}
{\mu_0\left(H(t_k)-H(t_{k-1})\right)}.
\label{eq:incremental_permeability}
\end{equation}

\begin{figure}[!htbp]
    \centering

    \begin{subfigure}{0.48\textwidth}
        \centering
        \includegraphics[width=\linewidth]{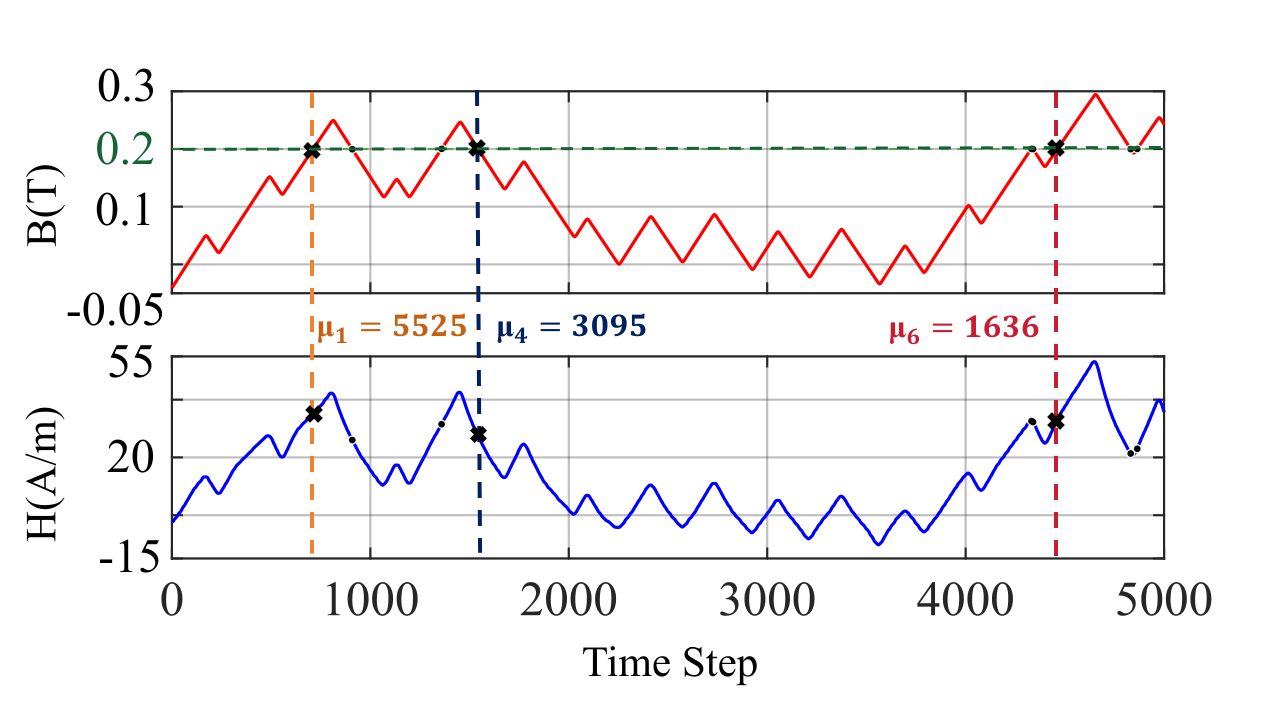}
        \caption{Incremental permeability at B = 0.2 T under 50kHz switching frequency}
        \label{fig:incremental_permeability}
    \end{subfigure}
    \hfill
    \begin{subfigure}{0.48\textwidth}
        \centering
        \includegraphics[width=0.6\linewidth]{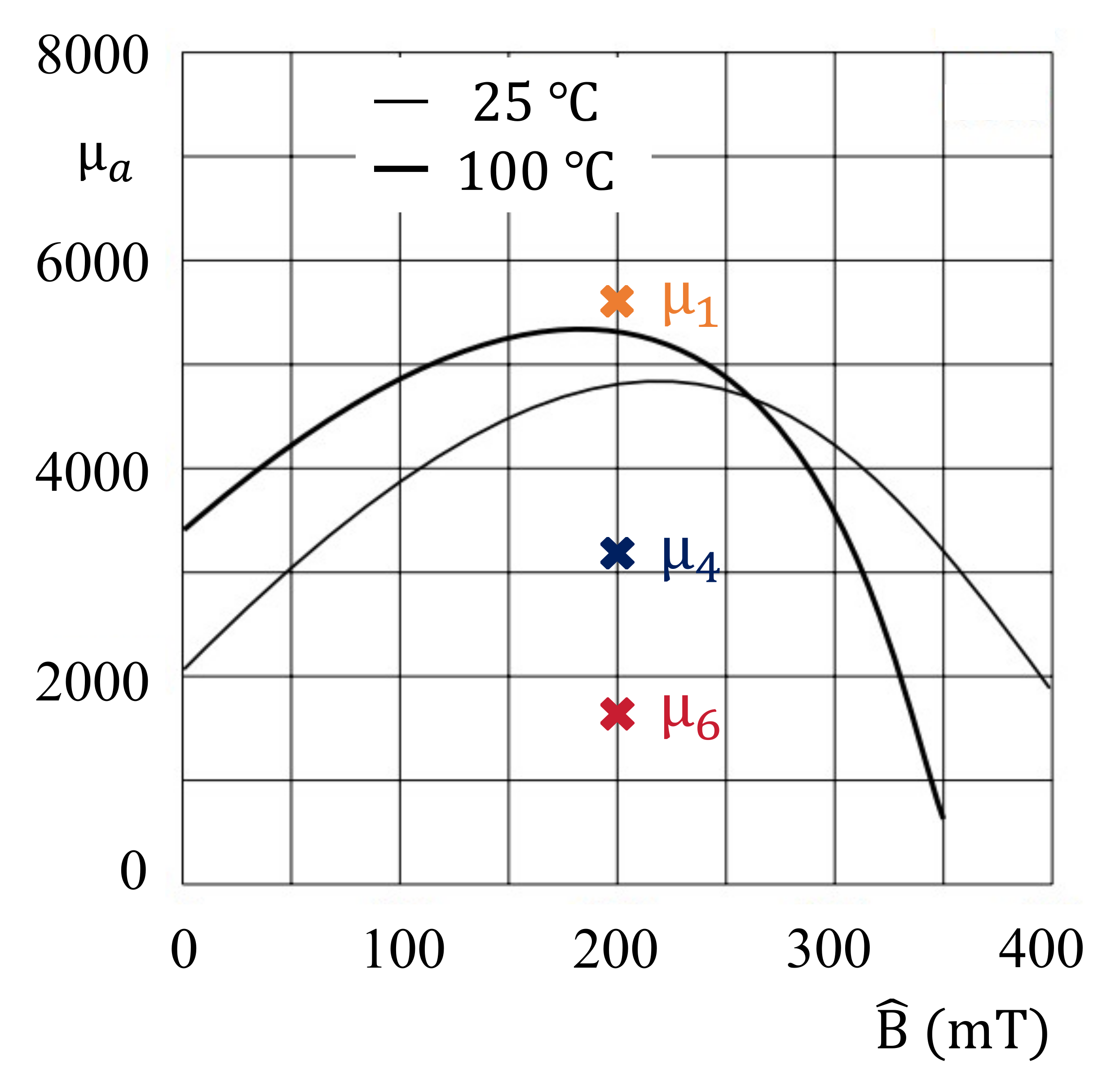}
        \caption{Mapping of the extracted incremental-permeability values on 3C90 datasheet amplitude-permeability [reprinted from \cite{ferroxcube3c90}]}
        \label{fig:datasheet}
    \end{subfigure}

    \caption{Variation of local incremental permeability under PWM excitation and qualitative comparison with 3C90 datasheet amplitude-permeability characteristic.}
    \label{fig:permeability_variation }
\end{figure}

The incremental permeability in practical operation can vary significantly
even under the same nominal state. Fig.~\ref{fig:incremental_permeability}
illustrates this behaviour using the measured $B$ and $H$ waveforms of 3C90
under a 50~kHz excitation. Taking $B=0.2~\mathrm{T}$ as an example, several
time instants exhibit the same instantaneous B(t) value but correspond to
different H(t) values, resulting in incremental-permeability values that
differ by more than a factor of three.

This dispersion arises because the incremental permeability depends not only
on the instantaneous magnitude of $B$, but also on the preceding $B$--$H$
trajectory, including the direction of traversal and recent reversals. Under
PWM excitation, each switching sub-cycle traces a minor loop, while the
aggregation of these minor loops forms the overall major quasi-static loop
\cite{wang2026magnetx}. The same instantaneous $B(t)$ value may therefore occur on different branches of the major loop or along different minor-loop trajectories,
reflecting different preceding magnetisation histories and resulting in different local $H(t)$ responses.. The incremental permeability is
further affected by operating conditions such as temperature, excitation
frequency, and DC bias.

For comparison with conventional material specifications, the
extracted incremental-permeability values $\Delta\mu$, are superimposed on the
manufacturer-provided amplitude-permeability characteristic in
Fig.~\ref{fig:datasheet}. The datasheet curve provides
a single amplitude-permeability value,
$\mu_a=\hat{B}/(\mu_0\hat{H})$, for each peak flux-density amplitude $\hat{B}$
under specified periodic test conditions. It is useful for material
characterisation but is not intended to represent the instantaneous local slope
of an evolving $B$--$H$ trajectory under PWM excitation. Although
$\Delta\mu$ and $\mu_a$ are not quantitatively equivalent, the comparison
highlights the limitation of representing transient magnetic behaviour using a
fixed permeability formula or a static single-valued $B$--$H$ mapping. Such
approaches cannot capture the continuously evolving minor loops and
history-dependent responses encountered under non-periodic PWM excitation,
thereby motivating a sequence-dependent modeling framework.

A data-driven neural-network model is a suitable method for addressing these
limitations. It can learn the complex nonlinear relationship between $B$ and
$H$ directly from measured $B$--$H$ trajectories, thereby reducing reliance
on a fully specified analytical description of the underlying magnetisation
mechanisms \cite{li2023magnet,antonio2021effective}. This approach is particularly suitable for
magnetic-material modeling because neural networks can capture the multidimensional dependencies associated with waveform shape, frequency, temperature, and magnetic history. However, a conventional feedforward neural network does not inherently retain information from preceding time steps \cite{sixdenier2008quasistatic}. Since the transient $H(t)$ response depends on both the instantaneous excitation and the preceding magnetic history, a recurrent architecture is required to preserve temporal memory and capture the history-dependent hysteresis dynamics.

Among recurrent architectures, we select LSTM for its gated memory mechanism, which retains relevant historical information over long sequences while suppressing less informative past observations \cite{gers2002learning}. The Seq2Seq framework is adopted because it can map a historical input sequence to a future output sequence while allowing the history and forecast windows to have different lengths. Therefore, a condition-aware LSTM-based Seq2Seq encoder--decoder architecture is developed, as described in the following subsection.

\section{Seq2Seq Framework and Modeling Pipeline}
The objective of transient $B$–$H$ modelling considered in this work is to reconstruct the unknown future $H(t)$ response from the available magnetic history and the known future $B(t)$ excitation. We adopt an LSTM-based sequence-to-sequence encoder-decoder strategy to reconstruct the corresponding $H(t)$ sequence from the magnetic history and known $B(t)$ trajectory, with temperature and effective PWM frequency included as conditional variables. The architecture allows the history and forecast windows to have different lengths. The following sections describe the model architecture and configurations, training and inference procedures, and saturation-aware auxiliary module.

\subsection{Model Architecture}
\subsubsection{LSTM-Based Seq2Seq Encoder-Decoder Structure}
\label{sec:LSTM_Structure}
LSTM is a recurrent neural network specifically designed to learn temporal dependencies in sequential data, making it particularly suitable for modeling transient processes where the current state depends on previous observations. Compared with conventional recurrent neural networks (RNNs), LSTM introduces gated memory cells that alleviate the vanishing-gradient problem and enable relevant historical information to be retained over long sequences \cite{hochreiter1997long}. Building on this capability, a sequence-to-sequence (Seq2Seq) formulation is adopted since transient magnetic modeling requires mapping a historical magnetic trajectory to a future response sequence rather than predicting a single scalar quantity. In this framework, the encoder transforms the historical sequence into a latent representation of the magnetic history, while the decoder uses this representation to generate the future response sequence, which is suitable for capturing the history-dependent dynamics and reconstructing the time-resolved magnetic response \cite{cho2014learning}.

\begin{figure*}[!t]
    \centering
    \includegraphics[
        width=\textwidth,
        height=0.35\textheight,
        keepaspectratio
    ]{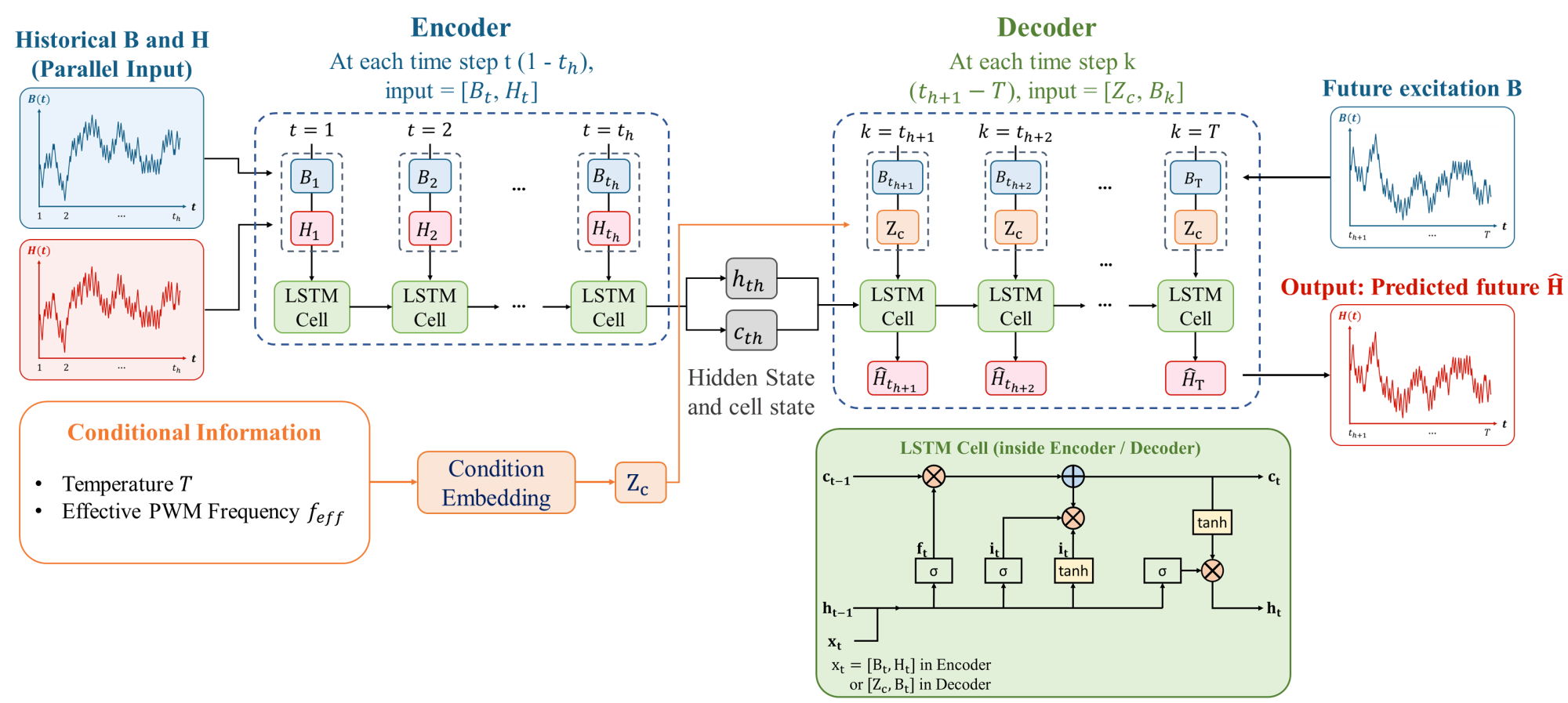}
    \caption{LSTM-based Seq2Seq encoder--decoder architecture with
    operating-condition embedding for transient $H$ prediction.}
    \label{fig:seq2seq_architecture}
\end{figure*}

As illustrated in Fig.~\ref{fig:seq2seq_architecture}, the proposed model constructs a
condition-aware LSTM-based Seq2Seq encoder--decoder architecture. At each historical
time step, the encoder receives the paired flux-density and magnetic-field inputs
$\left[B_t,H_t\right]$. It processes their temporal evolution recursively and compresses
the historical $B$--$H$ trajectory into the final hidden state $h_{t_h}$ and cell state
$c_{t_h}$. The hidden state represents the extracted historical features, while the
cell state preserves the longer-term memory accumulated over the historical sequence.
Both states are transferred to the decoder to initialise future sequence generation.
Since the encoder and decoder process the input and output sequences separately, the
two sequences are not required to have the same length
\cite{sutskever2014sequence}. This allows the model to accommodate different history
and forecast windows without modifying the underlying architecture.

In parallel, the operating conditions, including temperature and effective PWM frequency, are transformed into a latent condition vector $Z_{c}$ through a condition-embedding module. The historical $B$--$H$ trajectories and conditional information are processed through separate branches because they represent different forms of information. The LSTM encoder extracts history-dependent magnetic features from sequential waveforms, whereas the condition-embedding network maps sequence-level operating variables into a compact latent representation. This separation avoids repeatedly introducing static operating variables into the encoder and allows the temporal and conditional features to be learned independently before decoding.

During the forecast interval, the future flux-density excitation $B_k$ is known and supplied to the decoder, consistent with the voltage-driven operating conditions in most modern power converters, where $B$ is derived from applied voltage and $H$ is the unknown material response. At each forecast time step, the decoder combines the future excitation with the condition representation. Starting from the encoded state, it recurrently updates its hidden and cell states and produces the predicted magnetic-field value $\hat{H}_k$. Instead of predicting a single scalar quantity, the decoder reconstructs the complete time-resolved future $H$ waveform.

The figure also shows the internal structure of LSTM cell used in both encoder and decoder. At each time step, the LSTM updates its hidden state and cell state through gated operations, allowing relevant historical information to be retained while suppressing less informative content. The operation of each LSTM cell is expressed as:

\begin{equation}
\begin{aligned}
f_t &= \sigma(W_f x_t + U_f h_{t-1} + b_f), \\
i_t &= \sigma(W_i x_t + U_i h_{t-1} + b_i), \\
\tilde{c}_t &= \tanh(W_c x_t + U_c h_{t-1} + b_c), \\
c_t &= f_t \odot c_{t-1} + i_t \odot \tilde{c}_t, \\
o_t &= \sigma(W_o x_t + U_o h_{t-1} + b_o), \\
h_t &= o_t \odot \tanh(c_t),
\end{aligned}
\end{equation}

where $x_t$ denotes the input at time step $t$, $f_t$, $i_t$, and $o_t$ are the
forget, input, and output gates, respectively, $\tilde{c}_t$ is the candidate
cell state, and $c_t$ and $h_t$ denote the cell state and hidden state. The gates regulate the flow of information by controlling how much information is retained, updated or passed to the output at each time step. $\sigma(\cdot)$ denotes the sigmoid activation function, $\tanh(\cdot)$ denotes the hyperbolic tangent function, and $\odot$ represents element-wise (Hadamard) multiplication. The hidden state $h_t$ provides the output representation at time step $t$, while the cell state $c_t$ serves as the internal memory for preserving long-term temporal information. This memory capability is particularly relevant to dynamic $B$--$H$ modeling since the instantaneous H response depends not only on the current $B$ excitation but also on the preceding trajectory. By propagating an internal state across time steps, LSTM is well suited to representing the history-dependent magnetic behaviour. In the proposed framework,
$x_t = [B_t,H_t]$ for $t = 1,\ldots,t_h$ in the encoder and
$x_t = [Z_c,B_t]$ for $t = t_h+1,\ldots,T$ in the decoder.

The overall forward process of the proposed architecture is summarised in Algorithm~\ref{alg:seq2seq}.

\begin{algorithm}[!ht]
\caption{Condition-Aware Seq2Seq LSTM for Transient $H$ Prediction}
\label{alg:seq2seq}
\begin{algorithmic}[1]
\Require Flux-density sequence $\mathbf{B}$, historical magnetic-field sequence
$\mathbf{H}_{\mathrm{his}}$, operating condition vector $\mathbf{q}$,
history length $t_h$
\Ensure Predicted future magnetic-field sequence
$\widehat{\mathbf{H}}_{\mathrm{fut}}$

\State Split $\mathbf{B}$ into historical and forecast segments:
$\mathbf{B}_{\mathrm{his}}=\mathbf{B}_{1:t_h}$ and
$\mathbf{B}_{\mathrm{fut}}=\mathbf{B}_{t_h+1:T}$

\State Form the encoder input:
$\mathbf{X}_{\mathrm{enc}}
=
[\mathbf{B}_{\mathrm{his}},\mathbf{H}_{\mathrm{his}}]$

\State $(\mathbf{h}_{t_h},\mathbf{c}_{t_h})
\gets
\mathrm{LSTM}_{\mathrm{enc}}(\mathbf{X}_{\mathrm{enc}})$

\State $\mathbf{q}_{r}
\gets
\mathrm{ConditionReduction}(\mathbf{q})$

\State $\mathbf{z}_{c}
\gets
\mathrm{ConditionEmbedding}(\mathbf{q}_{r})$

\State Repeat $\mathbf{h}_{t_h}$ and $\mathbf{z}_{c}$ over the forecast length

\State Form the decoder input:
$\mathbf{X}_{\mathrm{dec}}
=
[\mathbf{h}_{t_h},\mathbf{z}_{c},\mathbf{B}_{\mathrm{fut}}]$

\State $\mathbf{Y}_{\mathrm{dec}}
\gets
\mathrm{LSTM}_{\mathrm{dec}}
(\mathbf{X}_{\mathrm{dec}},
(\mathbf{h}_{t_h},\mathbf{c}_{t_h}))$

\State $\widehat{\mathbf{H}}_{\mathrm{fut}}
\gets
\mathrm{Linear}(\mathbf{Y}_{\mathrm{dec}})$

\State \Return $\widehat{\mathbf{H}}_{\mathrm{fut}}$
\end{algorithmic}
\end{algorithm}

\subsubsection{Full-Window and Short-Window Configuration}
\begin{figure}[!htbp]
    \centering

    \begin{subfigure}{0.48\textwidth}
        \centering
        \includegraphics[width=\linewidth]{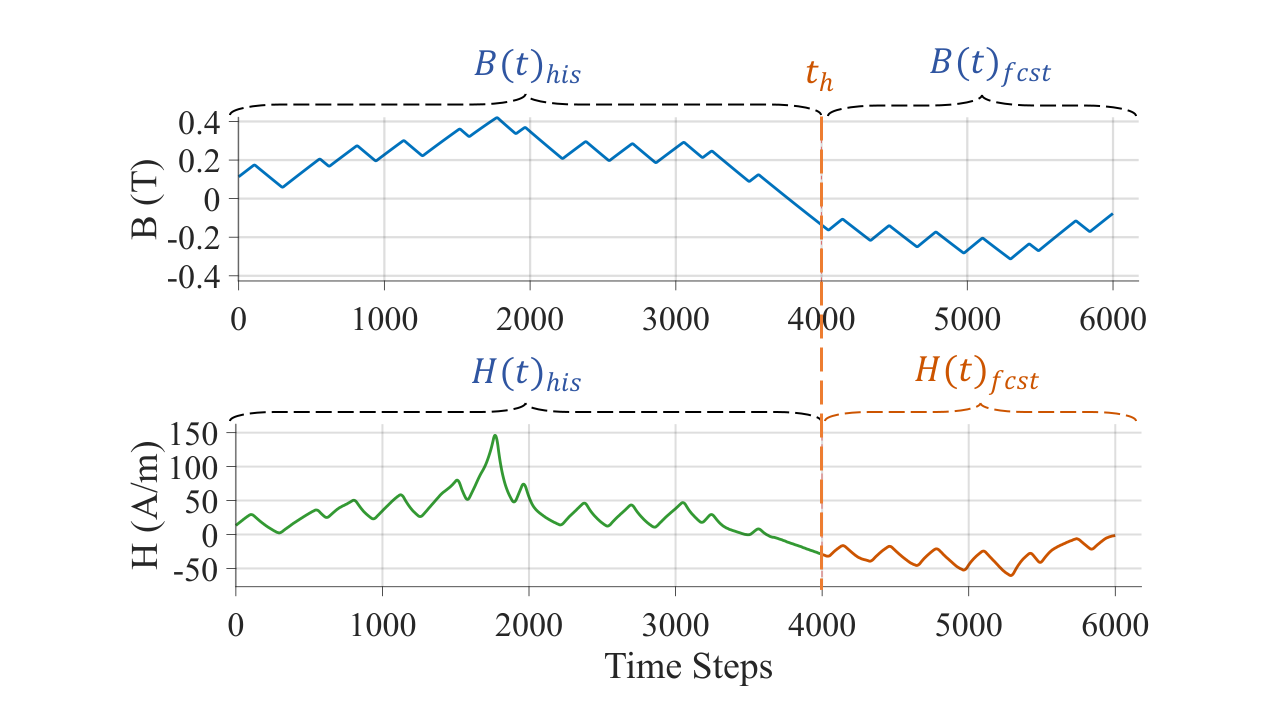}
        \caption{Full-window baseline with a fixed history–forecast partition}
        \label{fig:full_window}
    \end{subfigure}
    \hfill
    \begin{subfigure}{0.48\textwidth}
        \centering
        \includegraphics[width=\linewidth]{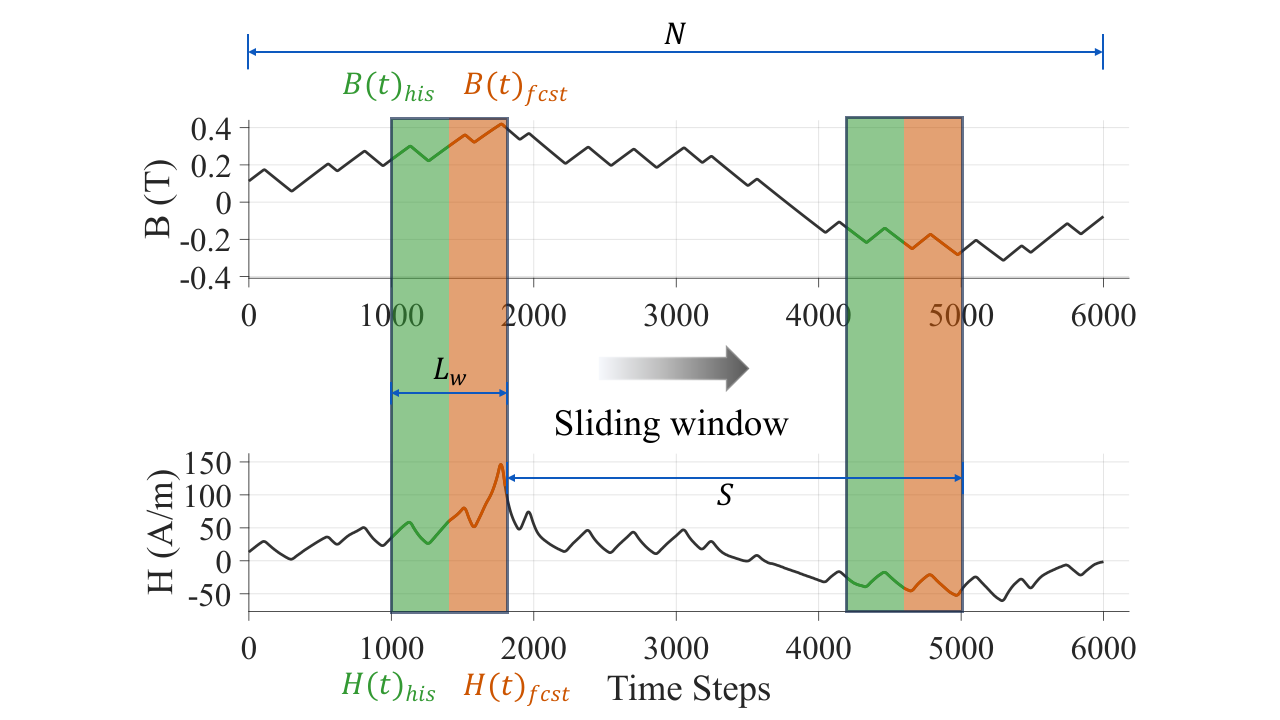}
        \caption{Short-window configuration for local sample extraction}
        \label{fig:short_window}
    \end{subfigure}

    \caption{Comparison of (a) full-window and (b) short-window sample construction for transient $B$--$H$ prediction.}
    \label{fig:window_configurations}
\end{figure}

\paragraph{Full-Window Configuration (Baseline)}
The full-window configuration is established as the reference setting. As illustrated
in Fig.~\ref{fig:full_window}, each fundamental-cycle sequence is
downsampled to a fixed length and divided at the history boundary $t_h$ into a
historical segment and a forecast segment. The paired historical flux-density and
magnetic-field sequences, $B(t)_{\mathrm{his}}$ and $H(t)_{\mathrm{his}}$, provide
the available magnetic memory. The subsequent flux-density sequence,
$B(t)_{\mathrm{fcst}}$, represents the known future excitation, while
$H(t)_{\mathrm{fcst}}$ is the corresponding response to be reconstructed. The
length-adaptive encoder--decoder architecture described in Section~\ref{sec:LSTM_Structure} can process
history and forecast windows of arbitrary lengths and generates an
$H(t)_{\mathrm{fcst}}$ sequence whose length matches that of the supplied
$B(t)_{\mathrm{fcst}}$.

This full-window baseline achieves a mean relative error of approximately 5.62\% across ten materials. However, varying the history-to-forecast ratio reveals that a longer historical segment does not necessarily lead to higher prediction accuracy. In some cases, reducing the history length decreases the relative error, indicating that distant historical observations may provide redundant information for the local transient response. Furthermore, treating each complete sequence as a single training sample limits the utilisation of local subsequences within the fundamental cycle. These observations motivate the short-window configuration introduced in the following subsection, which restricts the historical context and extracts multiple local samples from each full-length sequence.

\paragraph{Short-Window Configuration (Proposed)}
Motivated by the observation that distant historical information contributes
progressively less to local $H(t)$ prediction and that excessively long history
windows do not consistently improve accuracy, a short-window configuration is
developed using the same encoder--decoder backbone. Rather than processing the
complete sequence as a single sample, a local window of length $L_w$ is extracted
and divided into historical and forecast intervals, as shown in
Fig.~\ref{fig:short_window}. The window is shifted along the original
sequence using a stride $S$, generating multiple local forecasting samples from
each full-length trajectory.

By restricting the input window, this configuration focuses the encoder on the
recent magnetic history that is most relevant to each forecast segment. In
addition, sliding-window extraction acts as an effective data-augmentation
strategy by systematically traversing each full-length sequence, allowing the
data throughout the complete waveform to be utilised in multiple local training
samples. For a sequence containing $N$ samples, the number of extracted
subsequences is given by
\begin{equation}
A = \left\lfloor \frac{N-L_w}{S} \right\rfloor + 1,
\label{eq:augmentation_factor}
\end{equation}
where $A$ denotes the augmentation factor. The effects of the selected history lengths are quantitatively investigated in Section~\ref{sec:history_interpretability}.

\subsection{Data Preparation and Model Optimisation Strategy}

\subsubsection{Dataset Overview}
The  proposed work bases on the MagNet Challenge II dataset, which contains transient magnetic measurements covering a broad range of ferrite materials and operating conditions \cite{chen2026magnetchallenge2}. Ten materials are first used for base-model development and systematic evaluation of the proposed modeling framework. These materials provide relatively large datasets, with approximately 7,000–17,000 measured sequences available for each material. The developed modeling strategy is then extended to five previously unseen materials to evaluate material-specific adaptation and generalisation beyond the materials considered during base-model development. Overall, the dataset contains 148,214 measured sequences, with the sequence counts for individual materials summarised in Table \ref{tab:sequence_counts}.

\begin{table}[!htbp]
    \centering
    \caption{Sequence counts per material (MagNetX Database).}
    \label{tab:sequence_counts}

    \footnotesize
    \renewcommand{\arraystretch}{1.12}

    \begin{tabular}{c c}
        \toprule
        \textbf{Material} & \textbf{\# of sequences} \\
        \midrule

        \multicolumn{2}{l}{\textbf{Known-material (pre-training)}} \\
        3C90 & 17,269 \\
        3C94 & 10,413 \\
        3E6  & 10,583 \\
        3F4  & 11,100 \\
        77   & 11,909 \\
        78   & 10,726 \\
        N27  & 12,844 \\
        N30  & 12,105 \\
        N49  & 7,190  \\
        N87  & 12,617 \\

        \midrule

        \multicolumn{2}{l}{\textbf{Unseen-material (fine-tuning)}} \\
        3C92   & 105   \\
        3C95   & 6,758 \\
        FEC007 & 8,158 \\
        FEC014 & 9,536 \\
        T37    & 6,901 \\

        \midrule

        \textbf{Total} & \textbf{148,214} \\
        \bottomrule
    \end{tabular}
\end{table}

All waveforms are sampled at a fixed sampling frequency of 16 MHz, while the PWM switching frequency spans discrete operating points from 50 kHz to 800 kHz. For known materials, each recorded sequence contains 100 PWM switching sub-cycles. Consequently, the number of sampled time steps in each sequence varies inversely with switching frequency. The corresponding sequence lengths for all switching-frequency levels are summarised in Table \ref{tab:sequence_lengths}.

\begin{table}[!htbp]
\centering
\caption{Switching-frequency levels and typical sequence lengths at
$f_s=16$ MHz (100 PWM sub-cycles).}
\label{tab:sequence_lengths}
\renewcommand{\arraystretch}{1.12}
\setlength{\tabcolsep}{5pt}
\begin{tabular}{c c c c c c c c}
\hline
\textbf{$\bm{f}_{\mathrm{sw}}$ (kHz)}
& 50 & 80 & 125 & 200 & 320 & 500 & 800 \\
\hline
\textbf{$\bm{N}$ (samples)}
& 32016 & 20016 & 12816 & 8015 & 4975 & 3216 & 1967 \\
\hline
\end{tabular}
\end{table}

\subsubsection{Data Pre-Processing, Training, and Inference Workflow}
\begin{figure*}[!t]
    \centering
        \includegraphics[width=0.85\linewidth]{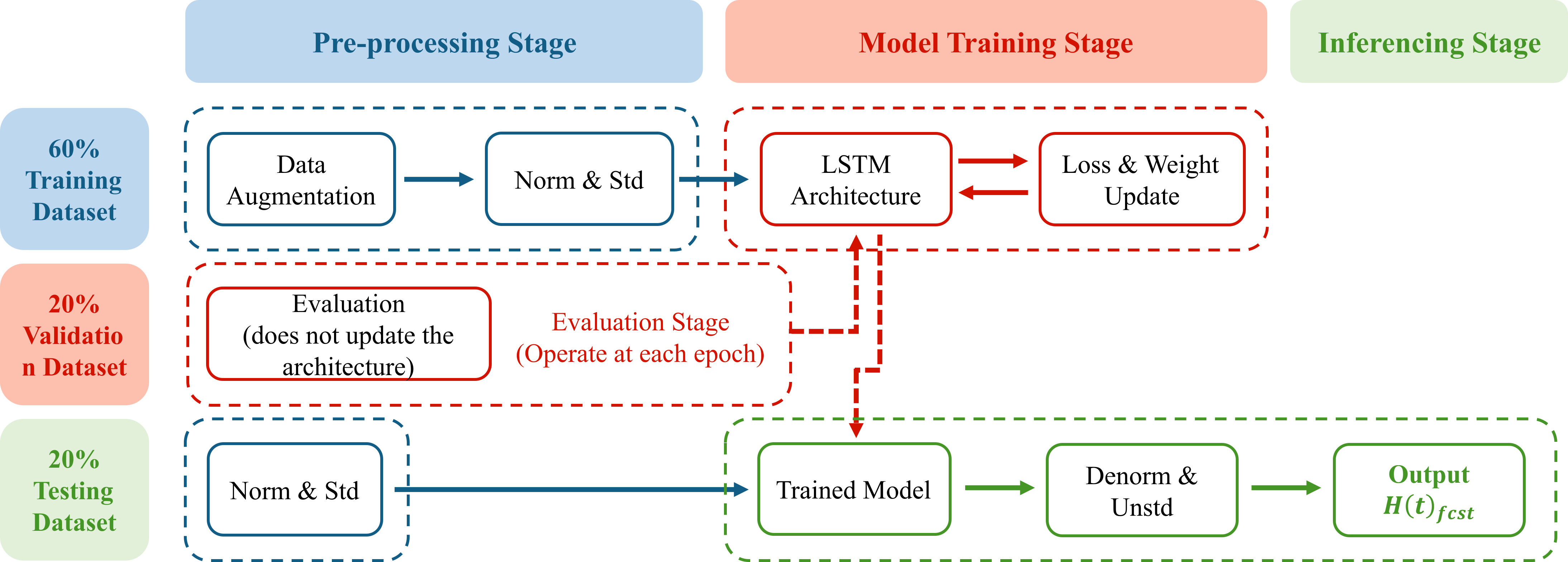}
        \caption{Overview of the model training and inference pipeline applied in base model construction stage.}
        \label{fig:pipeline}
\end{figure*}

The overall workflow of data pre-processing, model training, validation and inference is illustrated in Fig. \ref{fig:pipeline}. The original sequences are first divided at the waveform level into mutually exclusive training, validation, and testing subsets using a 60:20:20 ratio. This split is performed before subsequent processing to ensure that samples derived from the same measured trajectory cannot appear in different subsets, thereby preventing information leakage between model development and final evaluation. The training subset is used for model optimisation, while the validation subset is evaluated at each epoch without updating the model parameters. The testing subset is applied only for the final performance evaluation.

To accommodate the variable-length raw sequences while preserving their original temporal resolution, a fixed 1000-sample window is progressively shifted along each original sequence to extract local $B$--$H$ trajectory segments. Each extracted window is further constructed using three history-to-forecast configurations, namely 1:9, 5:5, and 9:1, representing different levels of magnetic history and prediction horizons. This configuration is relevant to practical transient modeling, where the available historical information and required forecast duration may vary with the operating scenario. It also allows the same length-adaptive Seq2Seq architecture to learn transient mappings over different prediction horizons rather than being restricted to a fixed history-to-forecast split. Sliding-window extraction simultaneously serves as data augmentation by generating multiple local samples from each measured trajectory and improving the utilisation of information distributed throughout the original sequence.

The extracted segments and conditional inputs are subsequently normalised using scaling parameters derived exclusively from the training subset, with the same transformations applied to the validation and testing subsets. During each epoch in the training stage, the same Seq2Seq model is sequentially optimised using samples from the three history-to-forecast configurations. For each batch, the predicted H waveform is compared with the corresponding target waveform using a 5\% quantile-trimmed mean-squared-error (MSE) loss, and the model parameters are updated through backpropagation using the Adam optimiser. The pointwise squared errors within each batch are combined, and values above the 95th percentile are excluded from the loss calculation. This trimming reduces the influence of rare high-error points (often associated with saturation events or under-represented patterns) while maintaining optimisation over the dominant inlier regime. Table \ref{tab:training_configuration} summarises the default hyperparameters used in the majority of training processes. At the end of each epoch, validation is performed independently for all three history configurations without gradient updates. The validation losses are then averaged with equal weighting. A model checkpoint, which stores the trained network weights and associated training state, is saved during training, and the checkpoint achieving the lowest combined validation loss is retained as the best model. During inference, each testing sample undergoes the same preprocessing before being passed to the trained model. The predicted H sequences are finally transformed back to their original physical scale and compared with the measured waveforms for performance evaluation.

\begin{table}[!t]
\centering
\caption{Default training configuration.}
\label{tab:training_configuration}
\renewcommand{\arraystretch}{1.12}

\begin{tabular}{cc}
\toprule
\textbf{Item} & \textbf{Value} \\
\midrule
Backbone        & Encoder--decoder LSTM (3 layers) \\
Hidden size     & 128 \\
Optimiser       & Adam \\
Learning rate   & $5 \times 10^{-4}$ \\
Batch size      & 128 \\
Loss            & MSE + 5\% quantile trimming \\
Window length   & 1000 samples \\
Checkpoint size & $\sim 9.88$ MB \\
\bottomrule
\end{tabular}
\end{table}

Model performance is evaluated on the testing subset using the sequence-level relative root-mean-square error (RMSE), defined as

\begin{equation}
\mathrm{RMSE}_{\mathrm{rel}}
=
\frac{
\mathrm{rms}\!\left(H_{\mathrm{pred}}(t)-H_{\mathrm{true}}(t)\right)
}{
\mathrm{rms}\!\left(H_{\mathrm{true}}(t)\right)
}.
\end{equation}

Before evaluation, the predicted H waveform is transformed back to its physical scale and aligned with the corresponding measured waveform. The testing results from the various history-to-forecast configurations are combined for overall evaluation. The mean relative RMSE across all testing sequences is used to characterise the overall prediction accuracy, while the 95th percentile of the sequence-level relative errors is reported as a complementary measure of performance on more challenging trajectories.

\subsubsection{Conditional Parameters: Effective PWM Frequency and Temperature with Ablation Analysis}
In addition to the time-dependent $B$--$H$ trajectory, temperature and PWM cycles are introduced as conditional parameters to account for variations in magnetic behaviour under different operating conditions. Temperature is directly included as a sequence-level condition, while the effective PWM-frequency descriptor is derived from the B(t) waveform. Significant positive and negative turning points are detected within an observation interval using a relative prominence criterion to suppress insignificant local fluctuations. The effective PWM-frequency descriptor is then defined as

\begin{equation}
f_{\mathrm{eff}}
=
\frac{N_{+}+N_{-}}{L_{\mathrm{int}}},
\end{equation}

where $N_{+}$ and $N_{-}$ denote the numbers of significant local maxima and minima detected in $B(t)$ waveform, respectively, and $L_{\mathrm{int}}$ is the number of sampled time steps within the observation interval. This descriptor characterises the density of switching activity in the corresponding sequence. It is not intended to recover the nominal switching frequency exactly; instead, it provides a consistent representation of the temporal switching characteristics that can be learned jointly with the magnetic response.

The observation interval used to construct the frequency descriptor depends on the amount of available history. When the historical interval contains sufficient switching information, the descriptor is evaluated directly from the historical $B$ segment. However, if the available history is too short to provide sufficient switching information, it can instead be evaluated from the complete available $B$ sequence, enabling the conditional information to remain meaningful in extreme cases where little or no H-history is available. Importantly, the same descriptor definition must be maintained during both training and inference, since the model learns the correspondence between the conditional input and the associated waveform dynamics rather than relying on its absolute numerical value alone.

\begin{table}[htbp]
    \centering
    \caption{Ablation study of temperature and effective PWM-frequency
    conditioning for transient $B$--$H$ prediction.}
    \label{tab:conditional_ablation}

    \footnotesize
    \renewcommand{\arraystretch}{1.12}

    \resizebox{\columnwidth}{!}{%
    \begin{tabular}{c c c}
        \toprule

        \begin{tabular}[c]{@{}c@{}}
        \textbf{Conditioning} \\
        \textbf{Configuration}
        \end{tabular}
        &
        \begin{tabular}[c]{@{}c@{}}
        \textbf{Mean Relative} \\
        \textbf{RMSE (\%)}
        \end{tabular}
        &
        \begin{tabular}[c]{@{}c@{}}
        \textbf{95th-Percentile} \\
        \textbf{Relative RMSE (\%)}
        \end{tabular}
        \\

        \midrule

        \begin{tabular}[c]{@{}c@{}}
        Temperature + Effective \\
        PWM Frequency
        \end{tabular}
        & 3.62 & 8.36 \\

        \arrayrulecolor{gray!35}
        \cmidrule(lr){1-3}
        \arrayrulecolor{black}

        Temperature Only
        & 5.40 & 13.44 \\

        \arrayrulecolor{gray!35}
        \cmidrule(lr){1-3}
        \arrayrulecolor{black}

        \begin{tabular}[c]{@{}c@{}}
        Effective PWM Frequency \\
        Only
        \end{tabular}
        & 3.74 & 9.26 \\

        \arrayrulecolor{gray!35}
        \cmidrule(lr){1-3}
        \arrayrulecolor{black}

        No Conditional Information
        & 4.47 & 11.16 \\

        \bottomrule
    \end{tabular}%
    }
\end{table}

Although switching characteristics are implicitly contained in the $B(t)$ waveform, it is not evident whether explicitly providing the effective PWM frequency contributes additional predictive information. An ablation study is therefore conducted using four conditioning configurations: both temperature and effective PWM frequency, temperature only, effective PWM frequency only, and no conditional parameters. The network architecture, training data and optimisation settings are kept unchanged across all configurations. As shown in Table~\ref{tab:conditional_ablation}, the configuration incorporating both temperature and effective PWM frequency achieves the lowest mean and 95th-percentile relative RMSE, confirming that the conditional parameters provide complementary information that improves prediction accuracy. A practical consideration is that both conditional parameters are normalised using scaling parameters derived from the training data. If either parameter during inference differs substantially from the training range, the model may operate outside its learned domain, potentially adversely affecting prediction accuracy.

\subsection{Saturation-aware Auxiliary Module}
Magnetic saturation represents a distinct nonlinear operating regime of ferromagnetic materials. Experimental characterisation of ferrite cores has shown that the magnetic permeability can decrease sharply as the operating point approaches saturation. For example, the relative permeability of a 3F3 ferrite decreases from a maximum of approximately 3900 to around 200 as the magnetic field increases towards saturation \cite{salas2013simple}. The pronounced changes motivate explicit consideration of saturation in transient modeling \cite{ma7031850}. Consistent with this material behaviour, evaluation of the baseline Seq2Seq model reveals that some of the most severe prediction failures occur when the forecast trajectory enters saturation region. In these regimes, the baseline model tends to produce an overly smooth H response and substantially underestimate the rapidly increasing field strength. 

\begin{figure*}[!t]
  \centering
  \includegraphics[width=\linewidth]{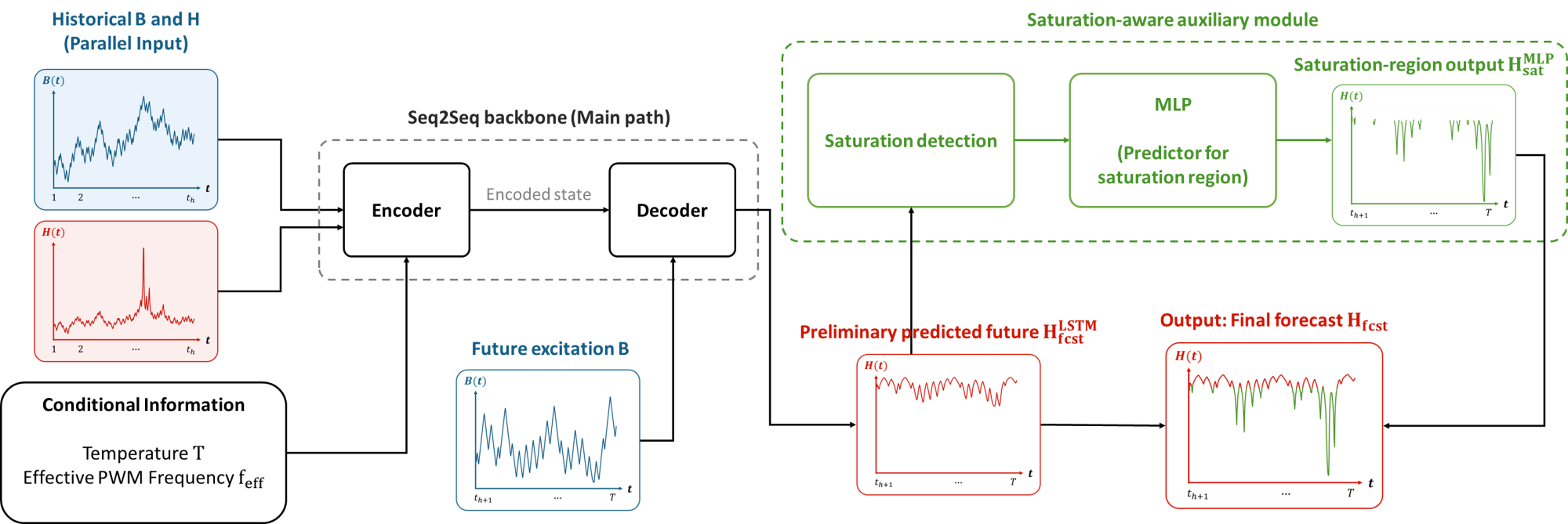}
  \caption{Architecture of the Seq2Seq backbone integrated with the saturation-aware auxiliary module.}
  \label{fig:pipeline_sat}
\end{figure*}

To improve robustness under saturation without perturbing the original Seq2Seq backbone, a saturation-aware auxiliary module is introduced as an additional prediction path, as illustrated in Fig \ref{fig:pipeline_sat}. Since the auxiliary branch operates only on the pre-identified saturated indices rather than modeling the complete history-dependent trajectory, a multilayer perceptron (MLP) a feed-forward neural network capable of nonlinear function approximation \cite{HORNIK1989359}, is adopted for the saturation region. The lightweight auxiliary MLP contains only 0.91\% as many parameters as the baseline LSTM model, resulting in negligible parameter overhead. It allows the relevant local relationship between the saturation-region excitation and H response to be learned without introducing an additional recurrent branch. A saturation threshold $H_{\mathrm{sat}}$ is applied to the preliminary sequence $H_{\mathrm{fcst}}(t)$ predicted by the main model to identify potentially saturated time steps. The saturation mask is defined as

\begin{equation}
m_k =
\begin{cases}
1, & \left|H_{\mathrm{fcst}}(t_k)\right| \ge H_{\mathrm{sat}} \\
0, & \text{otherwise}
\end{cases},
\label{eq:ml2_sat_mask}
\end{equation}
and the saturated inputs are extracted accordingly (e.g., $H_{\mathrm{fcst}}^{\mathrm{sat}}(t_k)=m_k\,H_{\mathrm{fcst}}(t_k)$). The saturated subset is fed into the MLP model that learns the specialised saturation mapping, while the original Seq2Seq structure remains unchanged and generate the complete forecast
$H_{\mathrm{fcst}}^{\mathrm{Seq2Seq}}(t)$ over the full interval. During inference, the predictions at the identified saturated time steps are replaced by the corresponding MLP outputs, while the remaining predictions are preserved,

\begin{equation}
H_{\mathrm{fcst}}^{\mathrm{final}}(t_k) =
m_k\, H_{\mathrm{sat}}^{\mathrm{MLP}}(t_k) + (1-m_k)\, H_{\mathrm{fcst}}^{\mathrm{Seq2Seq}}(t_k).
\label{eq:ml2_sat_fusion}
\end{equation}

\begin{figure}[!htbp]
  \centering
  \includegraphics[width=0.78\linewidth]{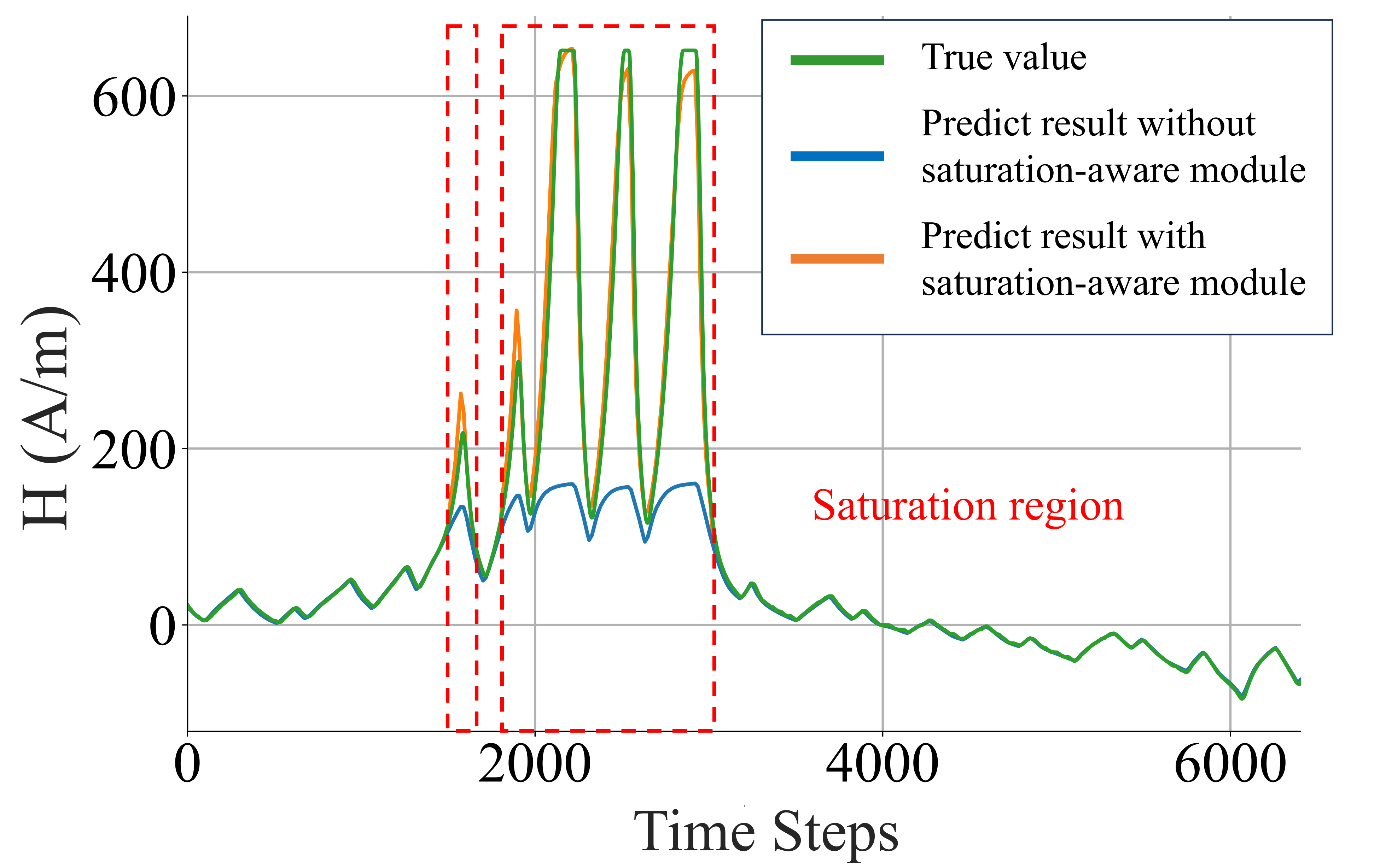}
  \caption{Representative 3C90 trajectory comparing the baseline Seq2Seq prediction and the saturation-aware prediction within the identified saturation region.}
  \label{fig:ml2_saturation_example}
\end{figure}

Fig. \ref{fig:ml2_saturation_example} presents the effect of the proposed correction using a representative 3C90 trajectory containing a saturation interval. In non-saturate region, the Seq2Seq prediction closely follows the measured H waveform. Once the trajectory enters the saturated region, the baseline forecast (blue) collapses into an overly smooth response and severely underestimates the peak amplitudes of $H(t)$, resulting in a sequence-level relative RMSE of 67.84\%. With the saturation-aware auxiliary branch enabled (orange), the prediction at saturated indices is substituted by the specialised MLP output, while the non-saturated portion remains governed by the original Seq2Seq branch. The resulting prediction reproduces the high-amplitude saturation peaks substantially more accurately and reduces the sequence-level relative RMSE to 9.19\%. Similar saturation-region improvements are observed for other evaluated materials, suggesting that the effectiveness of the proposed auxiliary module generalises across ferrite materials.

\section{Few-Shot Adaptation to Unseen Materials via Fine-Tuning}
\subsection{Representative Material-specific Industrial Scenarios}
To evaluate the generalizability of the LSTM-based Seq2Seq encoder-decode framework beyond the materials used for base-model development, five unseen target materials are introduced to represent different practical adaptation scenario. Rather than differing only in material identity, these datasets reflect limitations that may arise in engineering application, including restricted labelled data, incomplete operating-condition coverage, limited magnetic history, and material-level domain shift. The benchmark therefore provides a representative test of whether a model trained on data-rich reference materials can be efficiently adapted to new materials under incomplete measurement conditions.

Among the five materials, 3C92 is the ‘tiny data’ challenge, with only approximately 15\% of the training data available compared to other datasets. This reflects practical situations in which extensive waveform acquisition is expensive or impractical. 3C95 and FEC014 represent incomplete operating-condition coverage, with missing switching-frequency and flux-density regions, respectively. These cases require the model to generalise to operating points that are sparsely represented or absent during target-material adaptation. FEC007 represents an extreme limited history case, in which only a single historical H sample is available as an initial state before the future waveform must be reconstructed. Such a condition may arise when the model is initialized within a system-level simulation. This scenario provides a particularly challenging test because the lack of historical context effectively extends the prediction horizon and increases susceptibility to error accumulation with minimal prior magnetic-state information. T37 represents the cross-material out-of-distribution generalisation scenario, where the target material exhibits magnetic characteristics differ significantly from those presented during base-model development. 

The five-target materials probe complementary aspects of practical model transfer, ranging from data scarcity and operating-domain mismatch to missing historical information and material-level distribution shift. Training a separate model from scratch for each case would require sufficient target-specific data and repeated optimisation, which is impractical under several of these scenarios. Instead, the knowledge captured by the data-rich base models is reused as an initial representation and adapted to each target material through transfer learning and few-shot fine-tuning. The corresponding adaptation strategy is described in the following subsection.

\subsection{Few-Shot Adaptation Procedure}
\subsubsection{Base-Model Selection}
Transfer learning requires an appropriate source model whose learned magnetic representation is sufficiently compatible with the target material. Rather than selecting a base model according to material characteristic similarity, all ten models developed during the base-model construction stage are systematically evaluated as candidate initialisations for each unseen target material.

For each candidate, the corresponding pre-trained model (in checkpoint file) is loaded into the same Seq2Seq architecture and using a short adaptation run (e.g., 100 epochs) based only the training subset of the target material to avoid information leakage. Validation losses from all history-to-forecast configurations are averaged with equal weighting to form a combined selection criterion. The model achieving the lowest combined validation loss is retained for each candidate base model. The candidate with the best combined validation performance is then selected as the source model for subsequent material-specific fine-tuning. This data-driven selection strategy avoids relying on an assumed similarity between source and target materials and instead identifies the base model whose learned representation provides the most compatible initialisation across the different history-to-forecast conditions of the target material.

\subsubsection{Material-Specific Adaptation Strategies and Fine-Tuning Setup}
Fine-tuning is performed end-to-end from the selected base-model checkpoint, with all network layers remaining trainable. To preserve the transferable representation learned during base-model construction while allowing material-specific adaptation, Adam is used with a reduced learning rate, typically $1\times10^{-4}$ to $3\times10^{-4}$, and validation performance is monitored to mitigate overfitting.

Material-specific preprocessing is then introduced according to the characteristics of each target dataset. For the data-scarce 3C92 case, stronger augmentation is applied to increase the effective amount of supervision. In addition to the original waveforms, available historical segments with 500 and 900 samples are interpolated to 1000 samples and incorporated as additional training trajectories, while a smaller sliding stride of 100 samples is used to extract a denser set of local windows. FEC007 requires a dedicated treatment because only a single historical H sample is available. The conventional history-to-forecast ratio replication is therefore removed and the history length is fixed to one sample, corresponding to an extreme condition in which only the initial state is available. Since such a short historical interval cannot provide a meaningful estimate of the switching characteristics, the effective PWM-frequency descriptor is calculated from the complete $B$ sequence rather than from the historical segment alone. A ratio-matched base model is used before fine-tuning to reduce the discrepancy between the source training configuration and this extreme inference condition. For the remaining three materials, the standard protocol-matched preprocessing is retained, while the selected source model is instead fine-tuned end-to-end using the available target-material measurements.

Overall, the fine-tuning procedure provides a flexible transfer-learning framework capable of accommodating a range of practical engineering scenarios. By reusing the representations learned from data-rich source materials, the proposed approach enables adaptation using limited target-specific labelled data and avoids retraining a new model from scratch for each material. Since the same Seq2Seq backbone is retained across materials, the framework can be easily extended to new magnetic materials with only lightweight material-specific fine-tuning.

\section{Model-Size Optimisation and History-Length Analysis}
\subsection{Model Size-Accuracy Trade-off}
The predictive capability of an LSTM-based model is strongly influenced by its representational capacity, which is primarily determined by the hidden-state dimension and network depth. Increasing these parameters can improve the ability to learn complex temporal dependencies, but simultaneously increases the number of trainable parameters, memory requirements and computational cost. Previous studies compared LSTM architectures with different parameter counts and configurations, demonstrating that recurrent architectures should be evaluated not only by predictive performance but also by how effectively the available model parameters are utilised \cite{sak2014long}. Accordingly, the present study performs a controlled architecture-size analysis to identify a suitable balance between accuracy and model complexity for transient B-H modeling.

A series of Seq2Seq models with different hidden sizes and numbers of LSTM layers are independently trained using the same N87 dataset and training framework, with all other settings held constant. The investigated configurations range from H64-L1 to H192-L3, where H and L denote the hidden-state dimension and the number of LSTM layers, respectively. Table \ref{tab:model_size_analysis} summarises the resulting model sizes in terms of trainable parameter count, parameter memory, and stored checkpoint size. For each architecture, convergence is assessed using the combined training and validation behaviour across the different history-to-forecast configurations.

\begin{table*}[!htbp]
\centering
\caption{Model configurations and corresponding sizes for the size--accuracy analysis.}
\label{tab:model_size_analysis}
\renewcommand{\arraystretch}{1.12}
\setlength{\tabcolsep}{8pt}
\begin{tabular}{lccccc}
\hline
\textbf{Configuration} &
\textbf{Hidden Size} &
\textbf{LSTM Layers} &
\textbf{Trainable Parameters} &
\textbf{Parameter Size (MB)} &
\textbf{Checkpoint Size (MB)} \\
\hline
H64-L1  & 64  & 1 & 84,353    & 0.322 & 0.985  \\
H64-L2  & 64  & 2 & 150,913   & 0.576 & 1.755  \\
H96-L2  & 96  & 2 & 336,961   & 1.285 & 3.884  \\
H96-L3  & 96  & 3 & 485,953   & 1.854 & 5.596  \\
H128-L3 & 128 & 3 & 860,929   & 3.284 & 9.888  \\
H160-L3 & 160 & 3 & 1,342,401 & 5.121 & 15.398 \\
H192-L3 & 192 & 3 & 1,930,369 & 7.364 & 22.126 \\
\hline
\end{tabular}
\end{table*}

\begin{figure}[!htbp]
    \centering
        \includegraphics[width=0.95\linewidth]{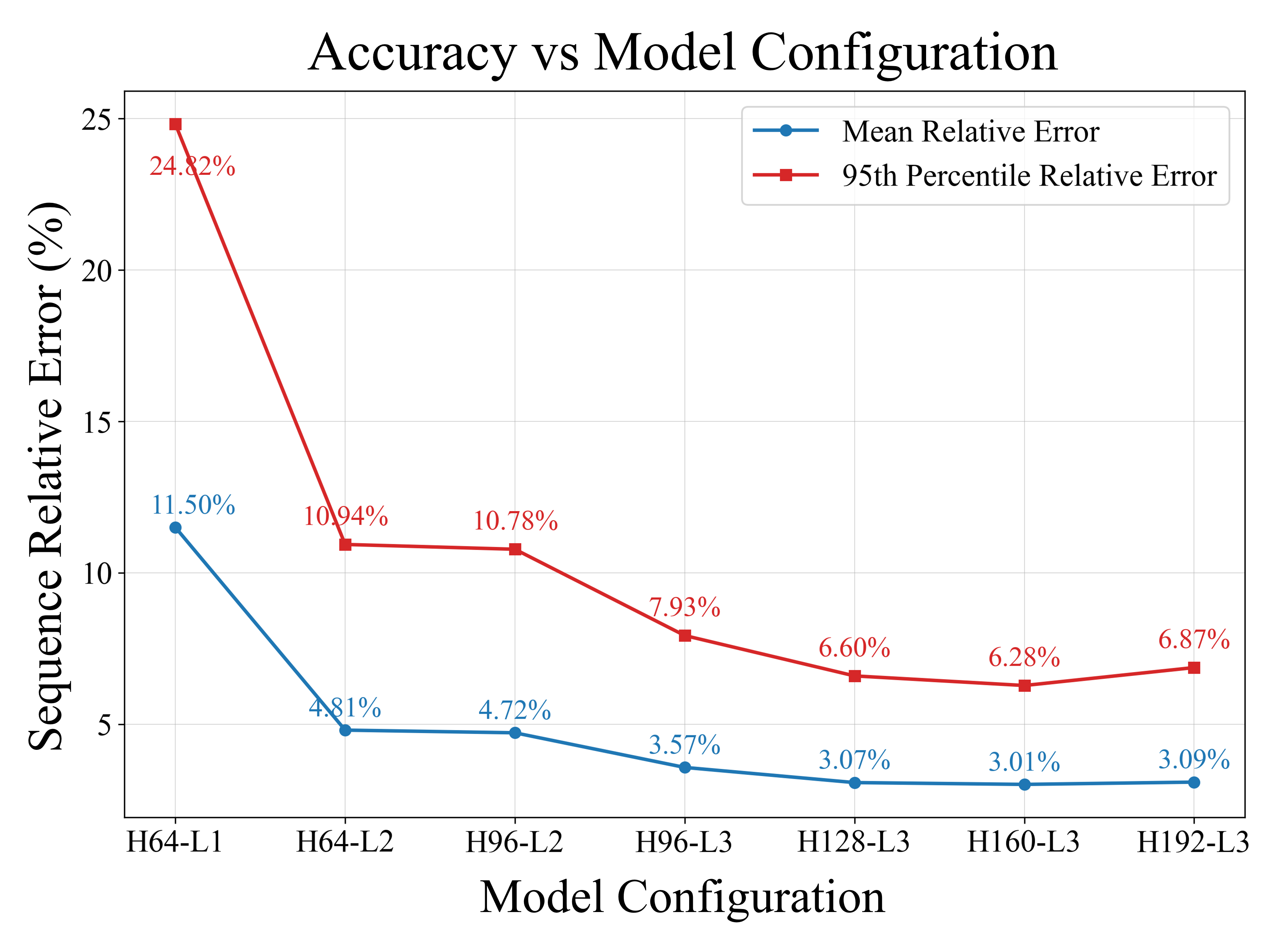}
        \caption{Prediction accuracy of different LSTM model configurations.}
        \label{fig:size_accuracy}
\end{figure}

Fig. \ref{fig:size_accuracy} compares the mean and 95th-percentile sequence-level relative errors obtained with different model configurations. Increasing the model capacity initially produces a substantial improvement in prediction accuracy, with H128-L3 achieving corresponding errors of 3.07\% and 6.60\%. However, the improvement becomes marginal relative to the increase in model size. Increasing the hidden-state dimension from 128 to 160 raises the trainable parameter count from 860,929 to 1,342,401, an increase of approximately 56\%, while reducing the mean relative error by only 0.06 percentage points and the 95th-percentile error from 6.60\% to 6.28\%. The H192-L3 configuration contains more than twice as many trainable parameters as H128-L3, while the corresponding prediction errors increase slightly to 3.09\% and 6.87\%. It also exhibits stronger transient fluctuations in both the training and validation losses before eventual convergence, suggesting greater optimisation instability with excessive model capacity.

Although H160-L3 model configuration achieves the lowest absolute prediction errors, its improvement over H128-L3 is small relative to the substantial increase in parameter count and memory requirement. H128-L3 therefore represents a favourable size-accuracy operating point and is selected as the reference architecture for subsequent training processes. These results demonstrate that increasing model capacity beyond the required complexity for the current task may provide diminishing accuracy returns and introduce unnecessary computational and memory overhead without corresponding improvements in generalisation.

\subsection{Model Interpretability: History-Length Sensitivity and Contribution Analysis}
\label{sec:history_interpretability}
Although magnetic hysteresis is inherently history-dependent, providing a longer historical sequence does not necessarily improve prediction performance. From the LSTM perspective, the encoder compresses the input history into finite hidden and cell states. Extending the history window with weakly relevant or drifting cycles may obscure the most predictive information and increase the risk of overfitting, leading to the degradation of generalisation \cite{pascanu2013difficulty,bahdanau2014neural}. Therefore, the complete history from the point of prediction to the beginning of fundamental cycle may not be necessary for the present task. This motivates a systematic investigation of the effect of history length on predictive performance, followed by an occlusion-based sensitivity analysis to identify the regions that contribute most strongly to the prediction.
\begin{figure}[!htbp]
    \centering
        \includegraphics[width=\linewidth]{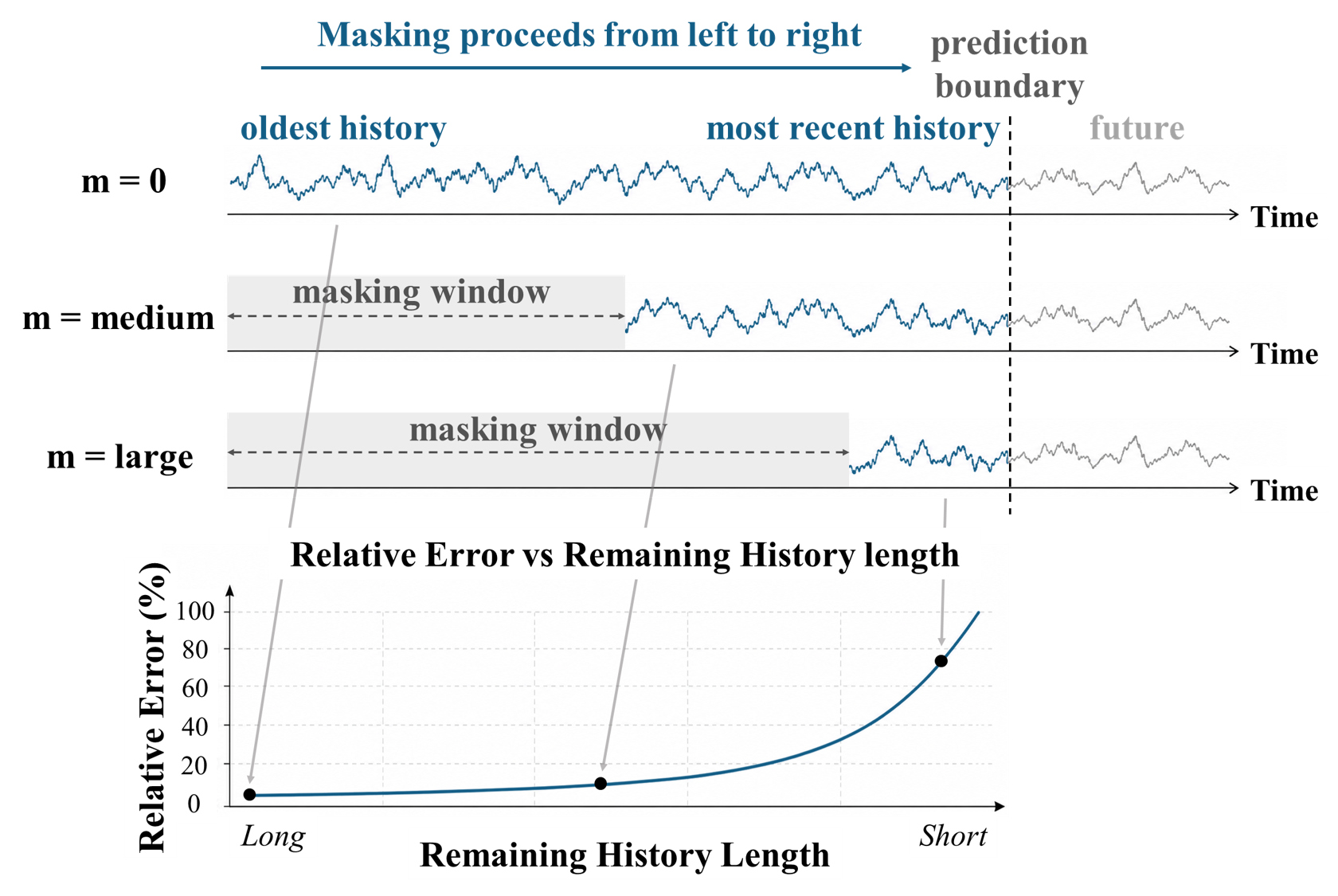}
        \caption{Illustration of progressive history occlusion for sensitivity analysis.}
        \label{fig:occlusion_illustration}
\end{figure}

Fig.~\ref{fig:occlusion_illustration} illustrates the process of the occlusion-based sensitivity study. Occlusion sensitivity is a perturbation-based interpretation method in which selected input regions are masked and their importance is assessed from resulting change in the model performance~\cite{zeiler2014visualizing}. Masking a subset of the input can be interpreted as replacing the selected components with a fixed value, thereby making the corresponding historical information unavailable to the model~\cite{vincent2010stacked}. This principle is adapted to the trained Seq2Seq LSTM to quantify the contribution of different historical segments without retraining a separate model for each history length~\cite{s21217373}. Since the prediction target is the future $H$ waveform, the analysis focuses on the H-history to quantify the model's sensitivity to the preceding magnetic response. As the synchronously sampled $B$ and $H$ sequences jointly describe the same magnetic trajectory, occluding either historical channel would follow the same general principle of removing historical trajectory information.

For each sample, the unmasked history ($m=0$) is first used to obtain the reference prediction and baseline relative error. The oldest $m$ points of the H-history are then replaced by a fixed masking value, while the remaining history, the complete $B$sequence and operating conditions are kept unchanged. The masked sample is passed through the model again, and its sequence-level relative error is calculated against the true future $H$ waveform. This procedure is repeated by progressively increasing $m$ until the entire H-history is masked. The resulting error variation is used to evaluate the effect of the available history length on prediction accuracy, as supported by previous studies relating LSTM prediction error to the number of preceding time steps~\cite{wielgosz2017using}.

\begin{figure*}[!t]
    \centering

    \begin{subfigure}[t]{0.49\textwidth}
        \centering
        \includegraphics[width=\linewidth]{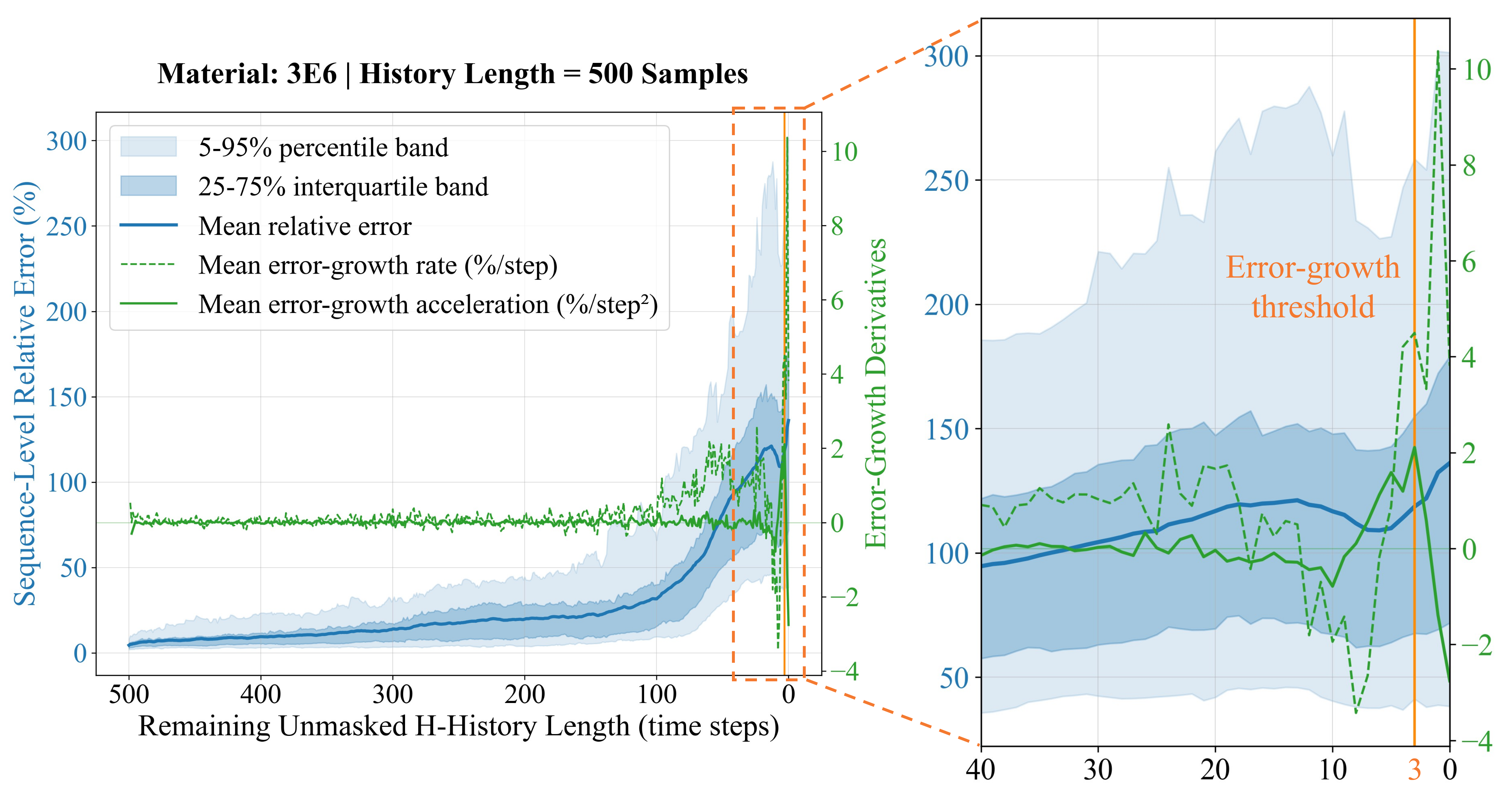}
        \caption{3E6 with 500 history samples}
        \label{fig:occ_3e6}
    \end{subfigure}
    \hfill
    \begin{subfigure}[t]{0.49\textwidth}
        \centering
        \includegraphics[width=\linewidth]{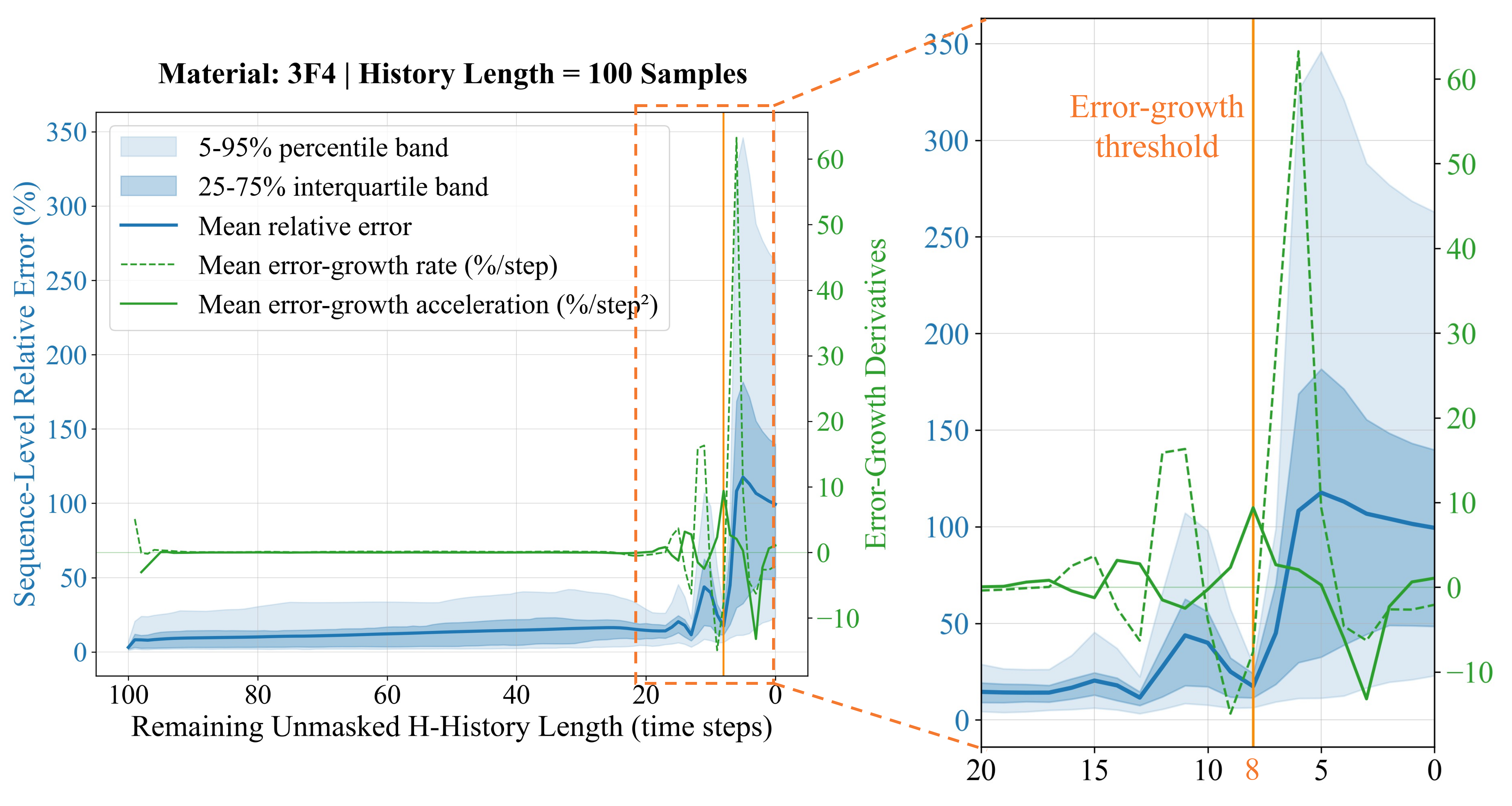}
        \caption{3F4 with 100 history samples}
        \label{fig:occ_3f4}
    \end{subfigure}

    \vspace{2mm}

    \begin{subfigure}[t]{0.49\textwidth}
        \centering
        \includegraphics[width=\linewidth]{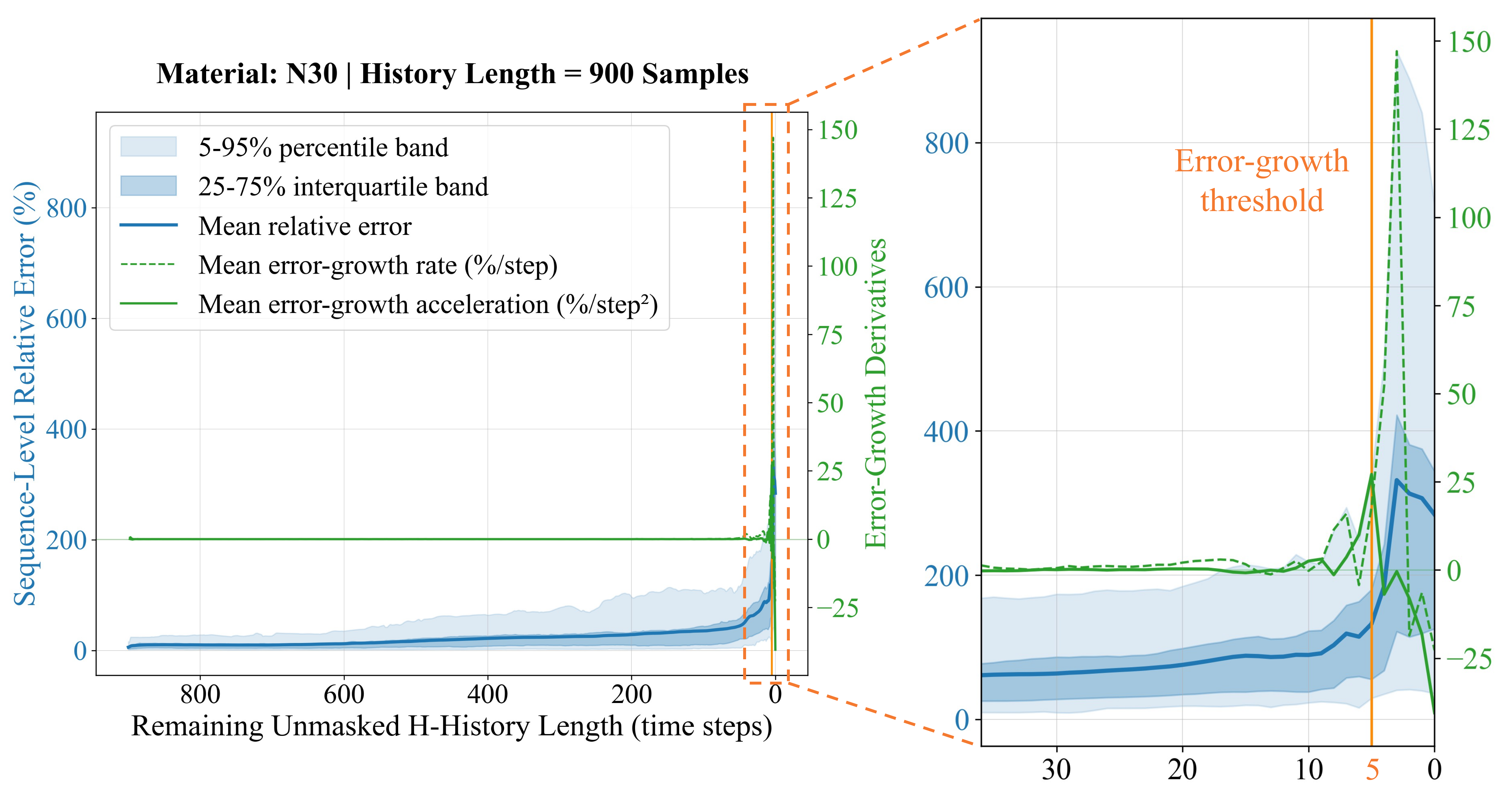}
        \caption{N30 with 900 history samples}
        \label{fig:occ_n30}
    \end{subfigure}
    \hfill
    \begin{subfigure}[t]{0.49\textwidth}
        \centering
        \includegraphics[width=\linewidth]{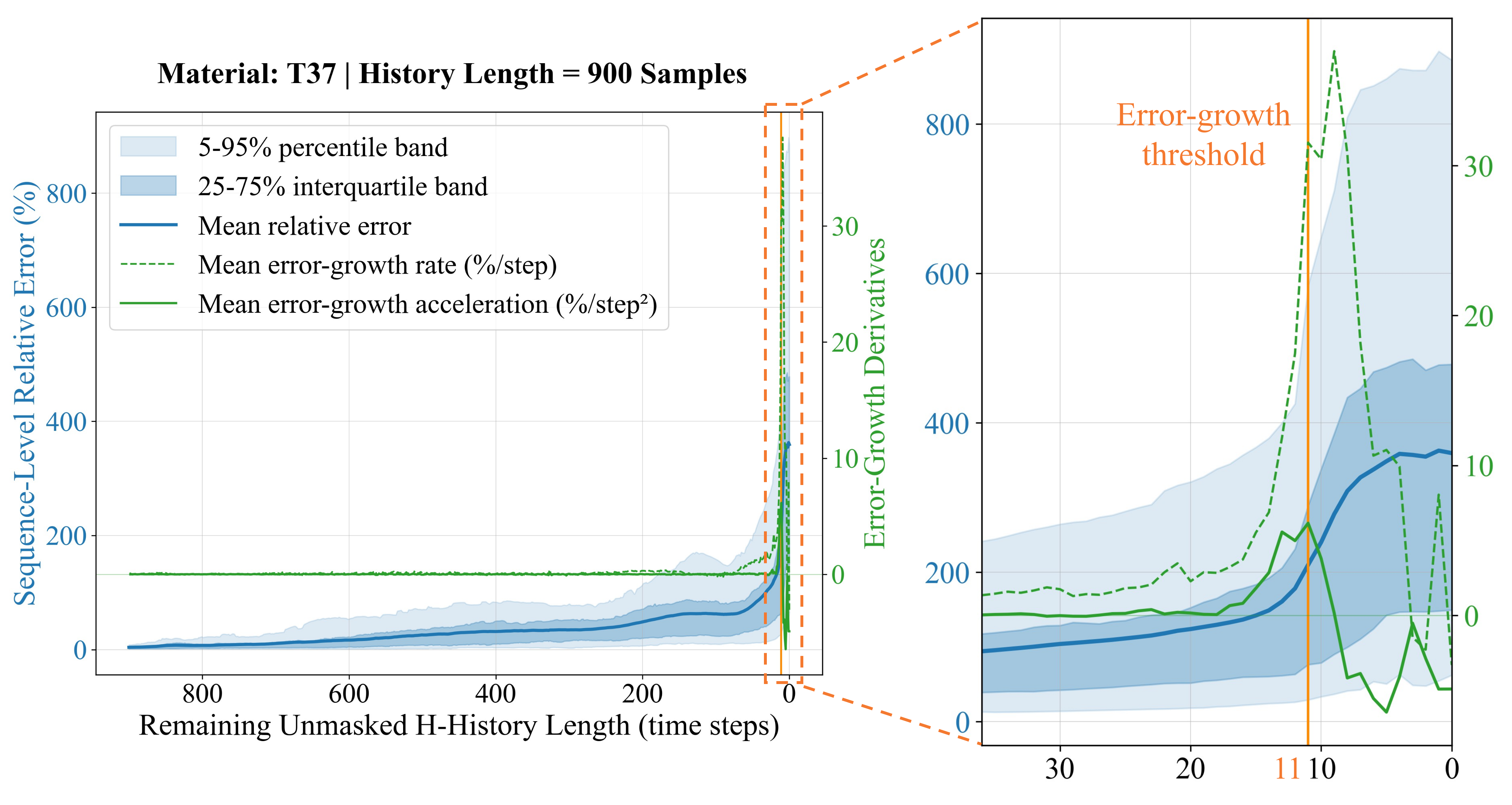}
        \caption{T37 with 900 history samples}
        \label{fig:occ_t37}
    \end{subfigure}

    \caption{Occlusion-based sensitivity analysis of the H-history for representative
    magnetic materials, with enlarged views showing the region of rapid prediction
    error growth and the identified error-growth threshold.}
    \label{fig:occlusion_results}
\end{figure*}

Fig.~\ref{fig:occlusion_results} shows representative occlusion-sensitivity results for different ferrite materials, total history lengths and operating conditions, including different switching frequencies and temperatures. The analysed samples are randomly selected from the original sequences and cover different history-to-prediction length ratios (e.g. 1:9, 5:5, 9:1). At each masking level, the sequence-level relative error is calculated as the RMSE between the predicted and measured future $H$ waveforms, normalised by the RMS value of the measured waveform. The blue solid curve represents the mean relative error across all selected samples at each remaining unmasked history length, while the dark- and light-blue shaded regions indicate the 25th--75th and 5th--95th percentile ranges, respectively. As the available history is progressively reduced, the prediction error generally increases gradually at first, followed by a rapid rise when only the most recent history is preserved.

To identify the point at which the prediction error begins to increase rapidly, the error-growth characteristics are further evaluated. The green dashed curve represents the mean error-growth rate, obtained from the first derivative of the relative-error curve for each sample, indicating the rate at which prediction error increases as historical information is removed. The green solid curve represents the corresponding mean error-growth acceleration, obtained from the second derivative, which characterises how rapidly the error-growth rate increases. The error-growth threshold, denoted by the orange vertical line, is defined at the maximum positive error-growth acceleration, indicating the onset of rapid deterioration in prediction accuracy. The enlarged views highlight this transition region and the identified threshold for each representative case.

\begin{figure}[!htbp]
    \centering
        \includegraphics[width=0.9\linewidth]{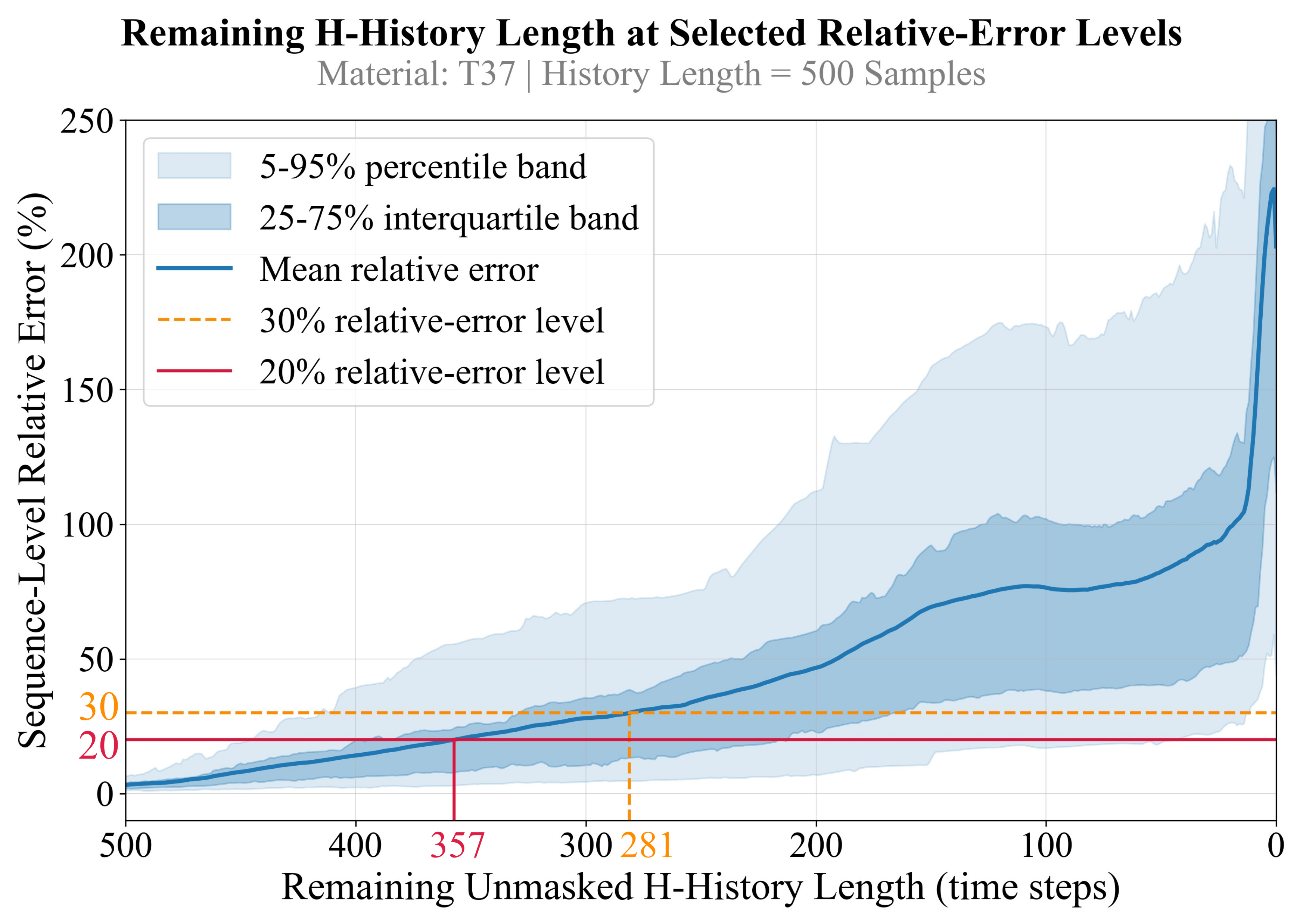}
        \caption{Remaining unmasked H-history lengths corresponding to selected sequence-level relative-error levels for T37 with a 500-sample history window.}
        \label{fig:occlusion_relerror}
\end{figure}

Fig.~\ref{fig:occlusion_relerror} further illustrates how the required H-history can be quantified for a prescribed prediction-error level, using T37 with a 500-sample history as a representative example. The horizontal lines indicate selected sequence-level relative-error levels. Their intersections with the mean relative-error curve determine the corresponding remaining unmasked H-history lengths, as presented by the vertical dashed lines. In this example, relative errors of 20\% and 30\% correspond to approximately 357 and 281 remaining history time steps, equivalent to 34.92~$\mu$s and 27.62~$\mu$s, respectively. From an engineering perspective, this relationship enables the history window to be selected according to a prescribed accuracy requirement: once an allowable prediction-error limit is specified for a particular application, the corresponding H-history provides an estimate of the historical information that should be retained to satisfy that requirement. This establishes a practical trade-off between prediction accuracy and history-window length, allowing unnecessary historical inputs to be removed and thereby reducing the computational and memory requirements during model deployment.

The same analysis is applied to all investigated materials, and the results are summarised in Table~\ref{tab:history_threshold}. The first two columns report the remaining H-history durations corresponding to mean relative-error levels of 20\% and 30\%, while the final column reports the error-growth threshold obtained from the occlusion-sensitivity analysis. The error-growth threshold is material dependent, but for most materials it occurs on the order of several hundred nanoseconds, with a median value of approximately 560~ns across the evaluated cases. Several materials exhibit longer history dependence, with thresholds extending into the microsecond range and reaching approximately 2.31~$\mu$s. The prescribed-error results show a similar material dependence, with a tighter accuracy requirement requiring a longer retained history. These results provide a practical basis for selecting the history window according to the allowable prediction error of a particular application.

\begin{table}[!htbp]
\centering
\caption{Error-growth thresholds and remaining H-history at selected relative-error levels for different magnetic materials.}
\label{tab:history_threshold}

\large
\renewcommand{\arraystretch}{1.12}

\resizebox{\columnwidth}{!}{%
\begin{tabular}{c c c c}
\hline
\textbf{Material} &
\begin{tabular}[c]{@{}c@{}}
\textbf{Error-Growth Threshold} \\
\textbf{for 500-Sample} \\
\textbf{History (ns)}
\end{tabular} &
\begin{tabular}[c]{@{}c@{}}
\textbf{Remaining H-History} \\
\textbf{at 20\% Relative Error} \\
\textbf{($\mu$s)}
\end{tabular} &
\begin{tabular}[c]{@{}c@{}}
\textbf{Remaining H-History} \\
\textbf{at 30\% Relative Error} \\
\textbf{($\mu$s)}
\end{tabular} \\
\hline
3C90   & 375    & 11.95 & 9.55  \\
3C94   & 500    & 12.55 & 10.86 \\
3E6    & 187.5  & 15.75 & 7.63  \\
3F4    & 500    & 2.19  & 0.99  \\
77     & 937.5  & 11.66 & 5.04  \\
78     & 187.5  & 11.66 & 3.40  \\
N27    & 687.5  & 43.06 & 14.66 \\
N30    & 312.5  & 27.28 & 11.62 \\
N49    & 812.5  & 12.86 & 5.55  \\
N87    & 2312.5 & 17.66 & 10.57 \\
3C95   & 1312.5 & 2.84  & 2.27  \\
FEC014 & 562.5  & 8.16  & 4.63  \\
T37    & 687.5  & 34.92 & 27.62 \\
\hline
\end{tabular}%
}
\end{table}

The results indicate that the historical information required for transient $H$ prediction is concentrated near the prediction boundary rather than distributed over the entire preceding trajectory. First, the complete history is not necessary, as a large proportion of the earlier H-history can be removed before a rapid deterioration in prediction accuracy is observed. The identified error-growth threshold also remains broadly consistent when different total history lengths are used, suggesting that it depends on the amount of recent informative history rather than on the initially available history window. Second, the relevant magnetic memory appears to be predominantly short-term rather than spanning a full fundamental cycle. Across the operating conditions represented in the MagNetX database \cite{chen2026magnetchallenge2}, the required history occupies only a small fraction of the corresponding fundamental period. Moreover, across the investigated switching and fundamental frequencies, the normalised error-growth threshold remains consistently close to the prediction boundary, with more than approximately 90\% of the preceding history removable in most cases before rapid error growth occurs. No clear dependence of this threshold on either frequency is observed. These findings suggest that, although the instantaneous magnetic response is inherently history dependent, the information required for predicting the subsequent $H$ waveform is primarily contained in the recent $B$--$H$ trajectory, which reflects the magnetic history immediately preceding prediction, rather than the complete switching- or fundamental-cycle history. Therefore, history dependence in transient magnetic behaviour does not necessarily imply long-term memory. For the present prediction task, effective magnetic memory is predominantly confined to a short time interval before the prediction boundary.

\section{Overall Performance and Generalisation}

\begin{table*}[!tb]
    \centering
    \caption{Overall prediction performance across the ten base-model materials and five unseen target materials, with comparisons against the constant-permeability baseline and the baseline in \cite{wang2025magnetx}.}
    \label{tab:overall_prediction_performance}

    \renewcommand{\arraystretch}{1.12}

    \resizebox{\textwidth}{!}{%
    \begin{tabular}{c c c c c c c}
    \hline

    \textbf{Material} &

    \begin{tabular}[c]{@{}c@{}}
    \textbf{Selected} \\
    \textbf{Base Model}
    \end{tabular} &

    \begin{tabular}[c]{@{}c@{}}
    \textbf{Mean Relative} \\
    \textbf{RMSE (\%)}
    \end{tabular} &

    \begin{tabular}[c]{@{}c@{}}
    \textbf{95th-Percentile} \\
    \textbf{Relative RMSE (\%)}
    \end{tabular} &

    \begin{tabular}[c]{@{}c@{}}
    \textbf{Constant-$\mu$ Baseline} \\
    \textbf{Mean Relative} \\
    \textbf{RMSE (\%)}
    \end{tabular} &

    \begin{tabular}[c]{@{}c@{}}
    \textbf{Constant-$\mu$ Baseline} \\
    \textbf{95th-Percentile} \\
    \textbf{Relative RMSE (\%)}
    \end{tabular} &

    \begin{tabular}[c]{@{}c@{}}
    \textbf{Baseline \cite{wang2025magnetx}} \\
    \textbf{95th-Percentile} \\
    \textbf{Relative RMSE (\%)}
    \end{tabular} \\

    \hline

    \multicolumn{7}{l}{\textbf{Base-model materials}} \\
    \hline

    3C90 & -- & 3.24 & 8.44  & 33.37  & 71.01  & -- \\
    3C94 & -- & 3.11 & 7.56  & 24.49  & 47.26  & -- \\
    3E6  & -- & 5.23 & 11.47 & 88.57  & 178.45 & -- \\
    3F4  & -- & 3.18 & 7.65  & 47.73  & 59.65  & -- \\
    77   & -- & 4.93 & 12.35 & 262.78 & 501.70 & -- \\
    78   & -- & 3.64 & 8.71  & 147.90 & 274.17 & -- \\
    N27  & -- & 2.75 & 6.04  & 113.43 & 219.12 & -- \\
    N30  & -- & 4.49 & 10.04 & 62.60  & 133.59 & -- \\
    N49  & -- & 3.82 & 8.60  & 53.09  & 141.13 & -- \\
    N87  & -- & 3.07 & 6.59  & 104.13 & 215.23 & -- \\

    \hline

    \multicolumn{7}{l}{\textbf{Unseen target materials}} \\
    \hline

    3C92   & N27 & 8.16 & 18.22 & 56.73 & 122.86 & 95.35 \\
    3C95   & N87 & 3.30 & 8.36  & 18.58 & 44.92 & 43.40 \\
    FEC007 & N87 & 5.79 & 13.29 & -- & -- & 96.95 \\
    FEC014 & N87 & 2.24 & 5.66  & -- & -- & 40.70 \\
    T37    & N27 & 3.95 & 8.61  & 58.13 & 142.57 & 57.84 \\

    \hline
    \end{tabular}%
    }

\end{table*}

Table~\ref{tab:overall_prediction_performance} summarises the overall prediction performance across the ten base-model materials and five unseen target materials, together with the comparisons against a datasheet-based constant-permeability baseline and the baseline in \cite{wang2025magnetx}. For the ten base-model materials, the mean relative RMSE ranges from 2.75\% to 5.23\%, while the corresponding 95th-percentile relative RMSE  remains between 6.04\% and 12.35\%. Averaged across these materials, the mean and 95th-percentile relative RMSE values are 3.75\% and 8.75\%, respectively, demonstrating consistently accurate prediction across different magnetic materials. 

In practical magnetic-component design, datasheet permeability is commonly used as an engineering parameter for inductance and magnetic-circuit calculations, with amplitude permeability $\mu_a$ specified for power ferrites under large-signal conditions \cite{ferrites1998user,serrano2023magnet}. Accordingly, a datasheet-based constant-permeability baseline is constructed using the manufacturer-specified $\mu_a$ where available, while initial permeability $\mu_i$ is applied otherwise. In each case, the selected permeability is treated as a fixed datasheet-derived reference value throughout the transient trajectory, giving $H(t)=B(t)/(\mu_0\mu_r)$, where $\mu_0$ is the vacuum permeability and $\mu_r$ is the selected relative permeability. Under this approximation, the prediction errors are substantially higher than those of the proposed framework, with the mean relative RMSE ranging from 24.49\% to 262.78\% and the 95th-percentile relative RMSE ranging from 47.26\% to 501.70\%. This comparison highlights the limitation of representing transient $B$--$H$ behaviour using a single fixed permeability, particularly when nonlinear and history-dependent magnetic responses are present.

Only two source models are selected for the five unseen targets: N87 is the preferred base model for 3C95, FEC007, and FEC014, whereas N27 is selected for 3C92 and T37. Since the corresponding target materials do not necessarily exhibit obvious similarities in material characteristics, this observation further supports the proposed performance-driven base-model selection strategy, in which source--target compatibility is determined through target-specific validation rather than assumed material similarity.

Following material-specific fine-tuning, the framework maintains high prediction accuracy across the five unseen target materials despite their different adaptation challenges. FEC014 achieves the lowest mean relative RMSE of 2.24\%, with a 95th-percentile error of 5.66\%, while 3C95 also achieves a relatively low mean error of 3.30\%. For the three unseen materials with available permeability data, the proposed model achieves consistently higher prediction accuracy than the corresponding constant-permeability baseline. Independent evaluation through MagNet Challenge II further demonstrates the competitiveness of the proposed framework \cite{chen2026magnetchallenge2}. It achieves substantially lower 95th-percentile relative RMSE than the Princeton benchmark across all five cases, with errors ranging from 5.66\% to 18.22\% compared with 40.70\% to 96.95\%.  Among the 32 participating teams, three of the five officially submitted models achieved top-three rankings, with FEC014 and T37 achieving first-place results and 3C92 ranking third. Although 3C92 exhibits the largest mean and 95th-percentile errors, at 8.16\% and 18.22\%, respectively, its third-place ranking suggests that these relatively high error is primarily associated with the difficulty of severely data-scarce scenario rather than reflecting a lack of model competitiveness. FEC007 represents an extremely limited-history case with minimal prior magnetic-state information. Despite this constraint, the model achieves a mean relative RMSE below 6\%, demonstrating that the Seq2Seq framework retains useful predictive capability even when historical $H$ information is almost entirely unavailable.

\begin{figure*}[!t]
    \centering

    \begin{subfigure}[t]{0.235\textwidth}
        \centering
        \includegraphics[width=\linewidth]{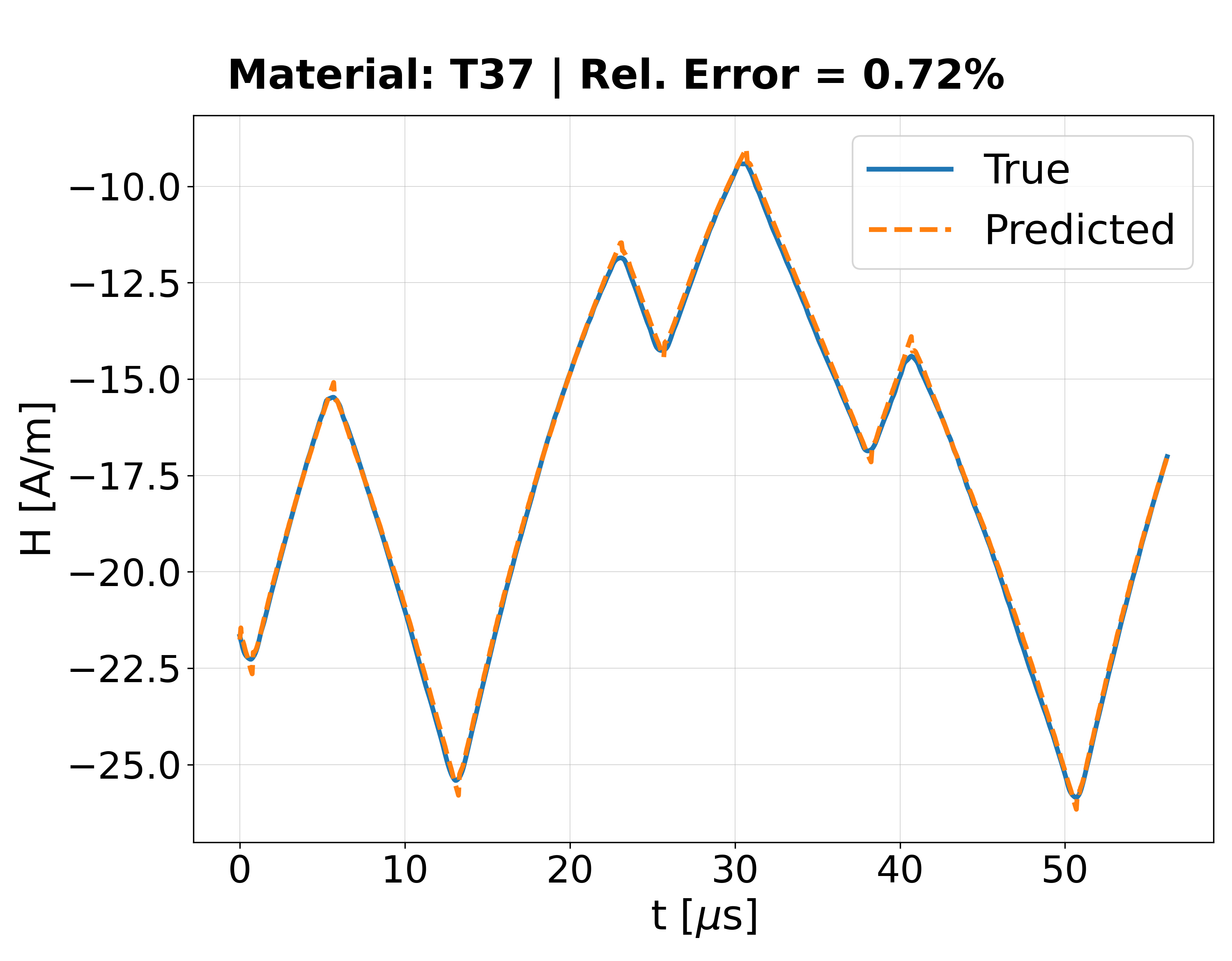}
        \caption{T37 waveform}
        \label{fig:t37_waveform}
    \end{subfigure}
    \hfill
    \begin{subfigure}[t]{0.235\textwidth}
        \centering
        \includegraphics[width=\linewidth]{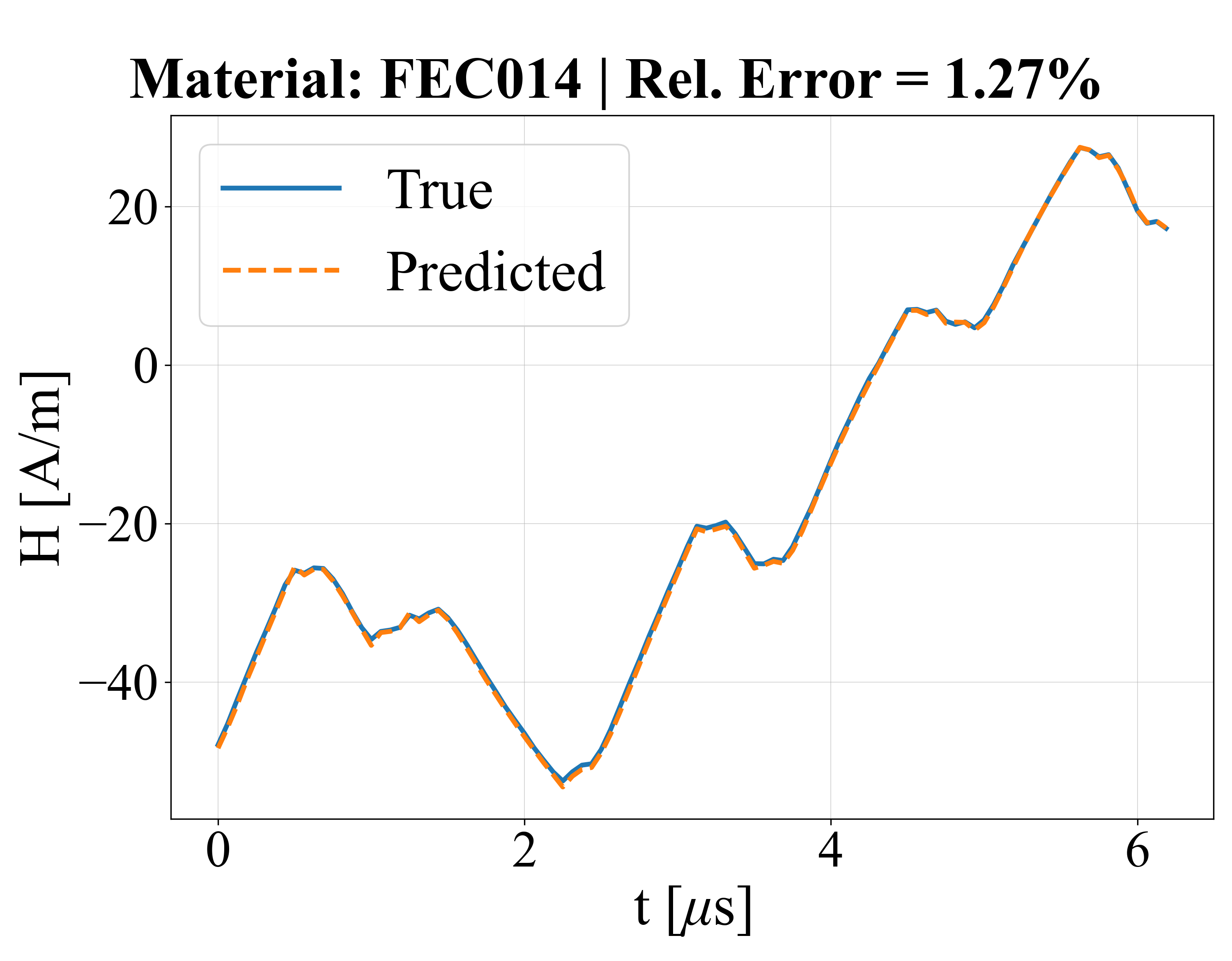}
        \caption{FEC014 waveform}
        \label{fig:fec014_waveform}
    \end{subfigure}
    \hfill
    \begin{subfigure}[t]{0.235\textwidth}
        \centering
        \includegraphics[width=\linewidth]{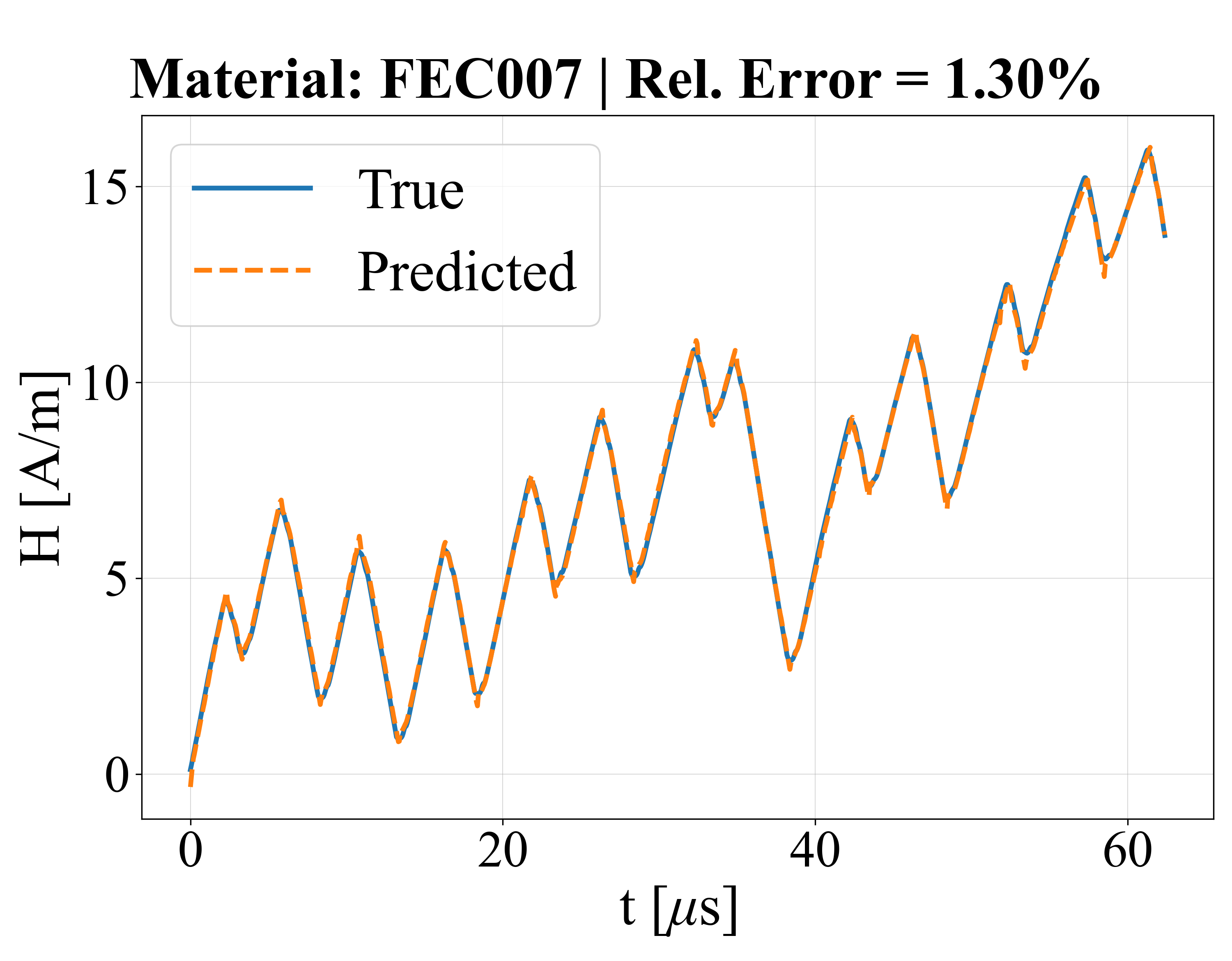}
        \caption{FEC007 waveform}
        \label{fig:fec007_waveform}
    \end{subfigure}
    \hfill
    \begin{subfigure}[t]{0.235\textwidth}
        \centering
        \includegraphics[width=\linewidth]{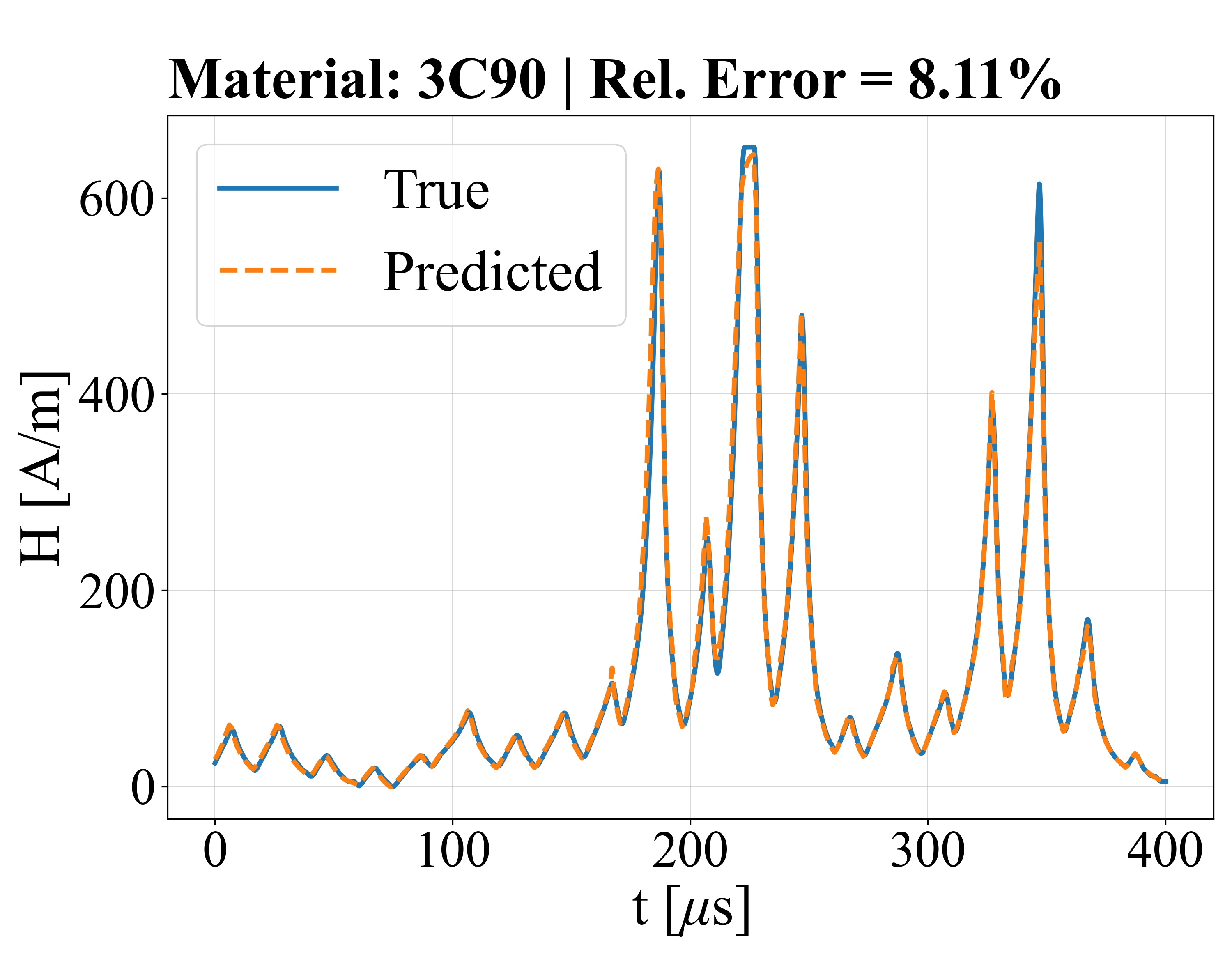}
        \caption{3C90 waveform}
        \label{fig:3c90_waveform}
    \end{subfigure}

    \vspace{2mm}

    \begin{subfigure}[t]{0.235\textwidth}
        \centering
        \includegraphics[width=\linewidth]{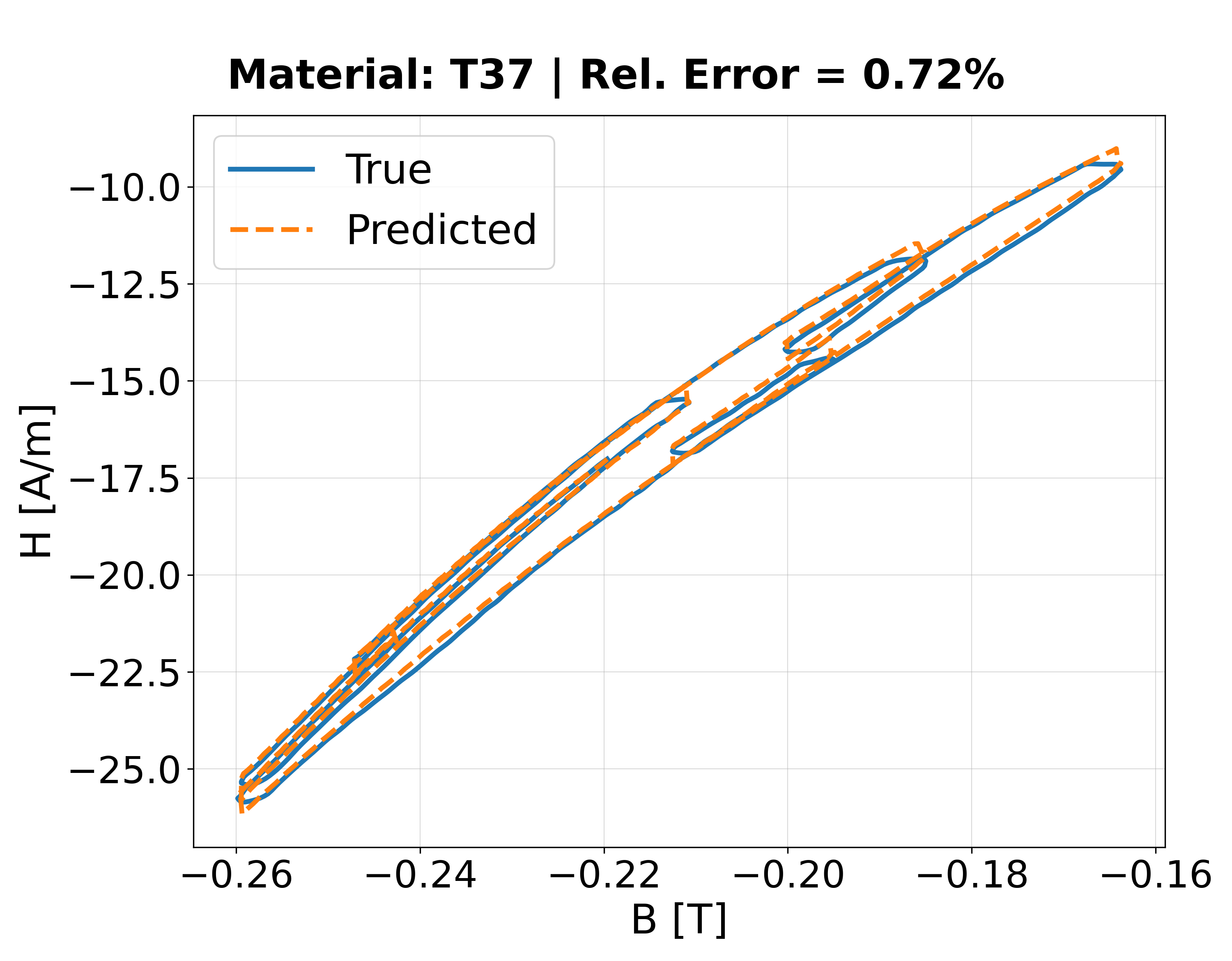}
        \caption{T37 $B$--$H$ trajectory}
        \label{fig:t37_bh}
    \end{subfigure}
    \hfill
    \begin{subfigure}[t]{0.235\textwidth}
        \centering
        \includegraphics[width=\linewidth]{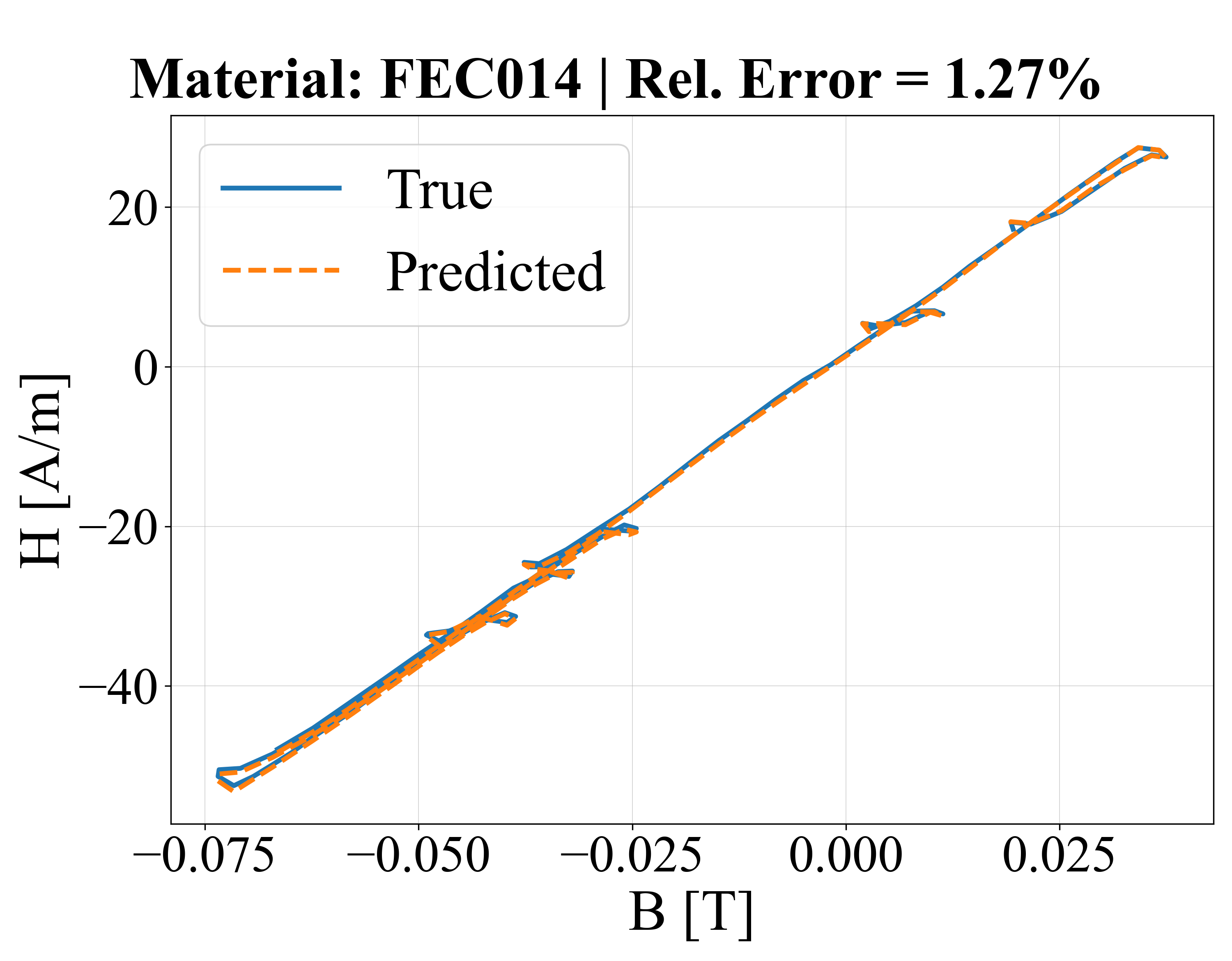}
        \caption{FEC014 $B$--$H$ trajectory}
        \label{fig:fec014_bh}
    \end{subfigure}
    \hfill
    \begin{subfigure}[t]{0.235\textwidth}
        \centering
        \includegraphics[width=\linewidth]{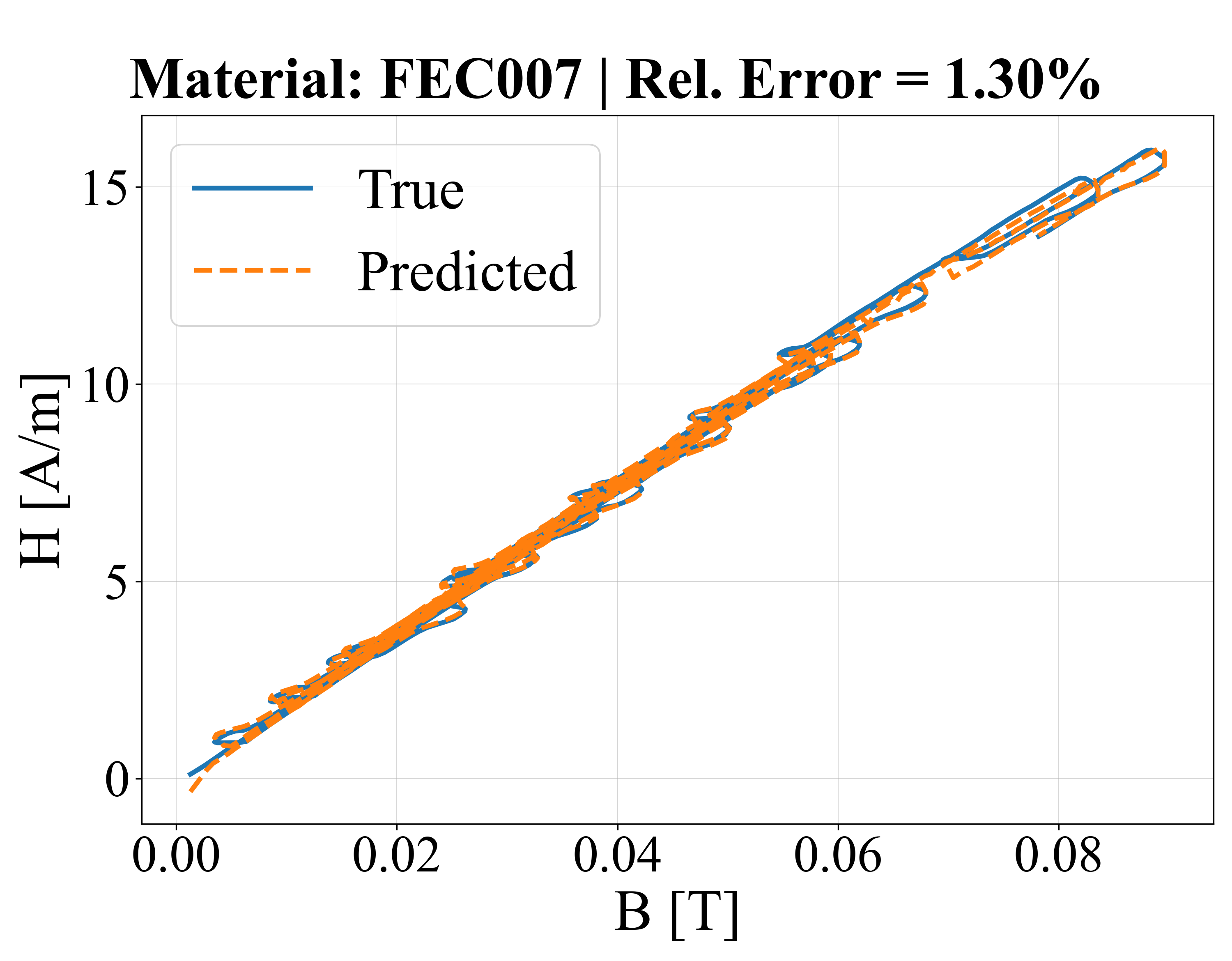}
        \caption{FEC007 $B$--$H$ trajectory}
        \label{fig:fec007_bh}
    \end{subfigure}
    \hfill
    \begin{subfigure}[t]{0.235\textwidth}
        \centering
        \includegraphics[width=\linewidth]{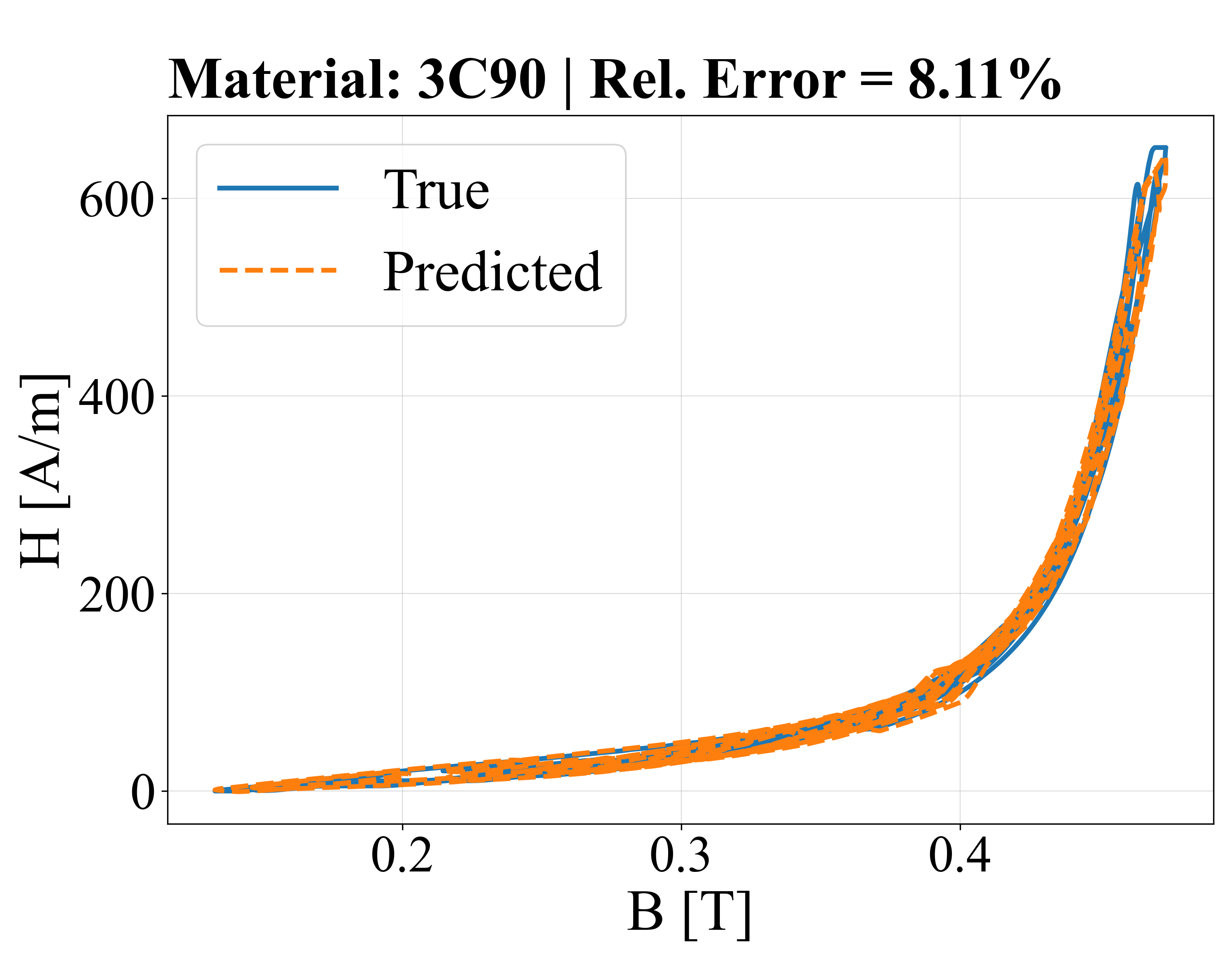}
        \caption{3C90 $B$--$H$ trajectory}
        \label{fig:3c90_bh}
    \end{subfigure}

    \caption{Comparison between the measured and predicted magnetic responses for representative materials: T37 for cross-material generalisation, FEC014 under extreme operating conditions, FEC007 with severely limited historical information and 3C90 in the saturation region. The first row shows the time-domain $H$ waveforms, while the second row shows the corresponding $B$--$H$ trajectories.}
    \label{fig:material_prediction_comparison}
\end{figure*}

Fig. \ref{fig:material_prediction_comparison} provides a qualitative assessment of the proposed framework under representative challenging operating conditions. The $H$ waveforms assess temporal prediction accuracy, while the corresponding $B$--$H$ trajectories verify the reconstruction of the underlying hysteretic behaviour. Fig. \ref{fig:t37_waveform} and \ref{fig:t37_bh} show the time-domain and hysteretic responses for T37, whose magnetic characteristics differ substantially from those of the base-model materials. Both results remain closely aligned with the measurements. Fig. \ref{fig:fec014_waveform} and \ref{fig:fec014_bh} evaluate FEC014 under an extreme operating condition combining the highest investigated switching frequency, approximately 800~kHz, and the highest temperature of $75\,^{\circ}\mathrm{C}$. Accurate reconstruction is maintained under this demanding combination of operating conditions. For FEC007, the close agreement between the measured and predicted $H$ responses in Fig. \ref{fig:fec007_waveform} and \ref{fig:fec007_bh} demonstrates that the model remains capable of reconstructing the subsequent magnetic dynamics under severely restricted historical information. Finally, Fig. \ref{fig:3c90_waveform} and \ref{fig:3c90_bh} illustrate a strongly saturated 3C90 case using the saturation-aware auxiliary module. Although $H$ increases sharply to several hundred A/m in the saturation region, the proposed framework successfully captures both the rapid waveform variation and the strongly nonlinear $B$--$H$ trajectory. Collectively, these representative cases demonstrate robust predictive fidelity across a broad range of challenging operating scenarios, including variations in material characteristics, extreme frequency and temperature conditions, severely limited historical information, and saturation behaviour.

Compared with next-step transient magnetic modeling approaches, the proposed Seq2Seq framework jointly encodes the paired historical $B$--$H$ trajectory and directly reconstructs the complete future $H$ sequence from the known future $B$ excitation. This formulation avoids explicitly feeding previously predicted $H$ values back into subsequent forecast steps, as adopted in single-step autoregressive approaches such as MagNetX \cite{wang2025magnetx}, thereby eliminating a recursive prediction-feedback mechanism through which multi-step errors may accumulate. This distinction is particularly relevant under severely limited-history conditions such as that represented by FEC007, where only minimal prior magnetic-state information is available. In addition, sequence-level operating conditions are processed through an independent condition pathway rather than being repeatedly introduced into the recurrent history encoder. The effective PWM-frequency descriptor further provides a compact representation of local PWM cycling and the associated minor-loops when the nominal switching frequency is not directly available.

The proposed framework also differs from physics-inspired hybrid approaches such as LSTM-DAO, which combines recurrent networks with predefined hysteresis operators \cite{wang2026framework}. In contrast, the Seq2Seq structure learns the transient input-response relationship directly from paired magnetic trajectories, while specialised mechanisms are introduced only where additional robustness is required. A dedicated auxiliary module for selectively correcting saturation-dominated predictions has not been reported in existing transient $B$--$H$ modeling studies. The proposed module provides targeted correction in saturation regions while preserving the non-saturated baseline prediction. Combined with the few-shot adaptation strategy, the pretrained model can be transferred from data-rich source materials to unseen target materials with limited labelled measurements. The framework therefore integrates direct future-sequence forecasting, condition awareness, saturation-specific correction, and cross-material adaptation within a unified architecture.

\section{Conclusion}
This paper presented an LSTM-based Seq2Seq encoder-decoder framework for time-resolved transient $B$--$H$ modeling under PWM excitation. The framework reconstructs the complete future $H$ waveform from the historical $B$--$H$ trajectory and known future $B$ excitation, with operating conditions to further increase prediction accuracy. A lightweight saturation-aware auxiliary module selectively improves robustness in saturation-dominated regions. Transfer learning with performance-driven base-model selection enables adaptation to unseen materials with limited data. Occlusion-based sensitivity analysis further shows that the most informative history is the recent local $B$--$H$ trajectory over only a few PWM cycles. History dependence therefore does not necessarily require fundamental-cycle-scale memory, providing both a physical interpretation and an engineering basis for selecting the retained history length. 

The framework is evaluated across 15 magnetic materials, achieving an overall mean relative RMSE of 4.06\% and a material-averaged 95th-percentile relative RMSE of 9.44\%. The deployment and evaluation code for  MagLearn~2 is publicly available on GitHub \cite{maglearn2_github}. The resulting waveform-level material model can potentially be integrated into converter simulation software to reproduce transient $B$--$H$ behaviour under realistic PWM excitation. Such integration could support more accurate component sizing and optimisation, reducing reliance on conservative design margins and thereby facilitating higher power density and more efficient magnetic components.

\bibliographystyle{IEEEtran}
\bibliography{references}

\end{document}